\documentclass[aps,prd,floats,floatfix,twocolumn,showpacs,eqsecnum,
superscriptaddress,nofootinbib]{revtex4-2}

\usepackage[centertags]{amsmath}
\usepackage{amssymb}
\usepackage{latexsym}
\usepackage{enumerate}
\usepackage{graphicx}
\usepackage{mathrsfs}
\usepackage{hyperref}
\hypersetup{
  colorlinks,
  citecolor={cyan!50!black},
  linkcolor={orange!60!black},
  urlcolor=violet,
}
\usepackage{stmaryrd}
\usepackage{import}
\usepackage{tensor}
\usepackage{mathtools}  %
\usepackage{commath}  %
\usepackage{physics}
\usepackage[usenames,dvipsnames]{xcolor}
\usepackage{bm}
\usepackage{multirow}
\usepackage{breakurl}
\usepackage{float}
\usepackage{verbatim}
\usepackage{mathrsfs}
\usepackage{booktabs}
\usepackage{orcidlink}
\usepackage{microtype}
\usepackage[capitalise]{cleveref}
\usepackage[caption=false]{subfig}
\usepackage[normalem]{ulem}
\crefname{section}{Sec.}{Sec.}
\Crefname{section}{Section}{Sections}
\newcommand{\ccite}[1]{Ref.~\cite{#1}}
\newcommand{\ccites}[1]{Refs.~\cite{#1}}

\newcommand{\defeq}{\coloneqq}

\newcommand{\dgF}{\mathcal{F}}
\newcommand{\dgFi}[1]{\tensor{\dgF}{_{\!{#1}}^i}}

\newcommand{\dgFA}{\mathcal{A}}

\newcommand{\bas}{\psi}
\newcommand{\lagr}{\ell}
\newcommand{\jac}{\mathrm{J}}
\newcommand{\invjac}{{(\jac^{-1})}}
\newcommand{\surf}[1]{{#1}^\Sigma}
\newcommand{\atp}[1]{\left.#1\right|}

\newcommand{\dgM}{M}
\newcommand{\dgD}{D}
\newcommand{\dgMD}{M\!\!D}

\newcommand{\dgML}{M\!\!L}
\newcommand{\dgMLD}{M\!\!L\!D}

\newcommand{\codename}[1]{\texttt{#1}}
\newcommand{\spectre}{\codename{SpECTRE}}
\newcommand{\vtu}{$vtu$}

\allowdisplaybreaks[1]

\definecolor{darkgreen}{rgb}{0.2,0.7,0.2}

\begin{document}

\title{Self-force calculations with numerical relativity methods}

\newcommand{\caltech}{\affiliation{Theoretical Astrophysics, Walter Burke
Institute for Theoretical Physics, California Institute of Technology, Pasadena,
California 91125, USA}}
\newcommand{\uzh}{\affiliation{Department of Astrophysics, University of Zurich, Winterthurerstrasse 190, 8057 Zurich, Switzerland}}
\newcommand{\aei}{\affiliation{Max Planck Institute for Gravitational Physics (Albert Einstein Institute), Am M\" uhlenberg 1, Potsdam 14476, Germany}}
\newcommand{\maryland}{\affiliation{Department of Physics, University of Maryland, College Park, Maryland 20742, USA}}
\newcommand{\geneseo}{\affiliation{Department of Physics and Astronomy, State University of New York at Geneseo, New York 14454, USA}}
\newcommand{\soton}{\affiliation{ School of Mathematical Sciences and STAG Research Centre, University of Southampton, Southampton, SO17 1BJ, United Kingdom}}
\newcommand{\ucd}{\affiliation{ School of Mathematics and Statistics, University College Dublin, Belfield, Dublin 4, Ireland}}
\newcommand{\cornell}{\affiliation{Cornell Center for Astrophysics and Planetary
Science, Cornell University, Ithaca, New York 14853, USA}}

\author{Nils L.\ Vu\,\orcidlink{0000-0002-5767-3949}} \email{nils.vu@uzh.ch} \caltech \uzh
\author{Nami Nishimura\,\orcidlink{0009-0002-4398-9130}} \aei \maryland
\author{Thomas Osburn\,\orcidlink{0000-0003-2747-3994}} \geneseo \ucd
\author{Jonathan E.\ Thompson\,\orcidlink{0000-0002-0419-5517}} \soton
\author{Lawrence E.\ Kidder\,\orcidlink{0000-0001-5392-7342}} \cornell
\author{Samuel D.\ Upton\,\orcidlink{0000-0003-2965-7674}} \soton
\author{Barry Wardell\,\orcidlink{0000-0001-6176-9006}} \ucd

\date{\today}

\begin{abstract}
  To model gravitational waveforms from extreme mass-ratio inspirals (EMRIs) for the upcoming LISA space mission, gravitational self-force calculations are needed to second order in perturbation theory.
  However, to date these calculations have only been attempted for the simplest case of circular orbits in Schwarzschild spacetime.
  In this work, we present a new computational method aimed at performing generic second-order self-force calculations in Kerr spacetime using methods from the adjacent field of numerical relativity. We perform an $m$-mode separation of variables, add null (``\vtu{}'') slicing in horizon-penetrating coordinates, and solve the resulting elliptic partial differential equations using high-order discontinuous Galerkin discretization, adaptive mesh refinement, and an iterative Krylov-type linear solver with parallelizable multigrid-Schwarz preconditioning.
  We find that our method achieves exponential convergence for the self-force on a scalar point charge in Kerr spacetime up to spins of $|a|/M=0.998$ (Thorne limit) on circular equatorial orbits as close as the innermost stable circular orbit (prograde and retrograde), despite the nonsmooth solution on the grid. We solve for 20 $m$-modes in parallel in a few seconds and retain the flexibility to extend the method to gravitational self-force and more generic orbits in the future.
  The code to perform these calculations is publicly available in the open-source numerical relativity code \spectre{}.
\end{abstract}

\maketitle

\section{Introduction}

Contemporary advancements in gravitational wave detectors such as the Laser Interferometer Space Antenna (LISA)~\cite{LISA,LISA:2024hlh,LISAWaveformWG:2023arg}, Einstein Telescope~\cite{ET:2025xjr} and Cosmic Explorer~\cite{Evans:2021gyd} will require development of new accurate compact binary waveform models with a redirected focus to meet upcoming observational needs. One such need involves extreme mass-ratio inspiral (EMRI) modeling; the lower-frequency gravitational waves emitted by EMRIs will be within the peak sensitivity range of LISA and will enable unprecedented tests of general relativity~\cite{EMRI}. The LIGO-Virgo-KAGRA era of compact binary modeling has largely focused on comparable mass systems because the associated frequency sensitivity is optimized for stellar mass binary components~\cite{Abbott_2019,Abbott_2021,Abbott_2022}. This environment motivated significant investment in numerical relativity calculations~\cite{Scheel:2025jct, Healy:2022wdn, Jani:2016wkt, Ferguson:2023vta, BAMcatalog}, which excel in the comparable mass regime but suffer from a catastrophic separation of scales when applied to EMRIs~\cite{LISAWaveformWG:2023arg}. Similarly elevated investments into EMRI modeling are now being motivated by the impending LISA mission.

The natural approach for asymmetric-mass binary modeling involves black hole perturbation theory and gravitational self-force calculations~\cite{Mino_1997,Quinn_1997,Poisson:2011nh,Barack:2018yvs,Pound_2021}; this strategy leverages the disparate masses by introducing a small mass-ratio parameter which is used to perturbatively expand the field equations and equations of motion. Accurate EMRI models will require the metric perturbation expansion to be carried out through second order in the mass ratio~\cite{Pound_2005,Burke_2024}. There are many choices to be made during formulation of the perturbation equations; different options include frequency domain vs.\ time domain, full separation of variables into $lm$-modes (separate both $\varphi$ and $\theta$) vs.\ partial separation into $m$-modes (separate only $\varphi$), and which gauge to adopt. Recent work in the frequency domain with $lm$-modes and Lorenz gauge led to successfully calculating second-order perturbations of a Schwarzschild black hole (with associated waveform templates) for the first time~\cite{Pound_2020,Wardell_2023,Albertini_2022,vandemeent_2023}.

The next step in this endeavor is the pursuit of second-order Kerr perturbations. This introduces significant new complexity, both because the orbital configurations can be more complicated (including precession effects, for example) and because the reduced symmetry of Kerr spacetime makes the perturbation equations less amenable to a separation-of-variables treatment. One of the most challenging problems faced by all second-order calculations is the construction of the second-order source, which is a quadratic function of the first-order metric perturbation and its first and second derivatives. Existing Schwarzschild calculations exploited spherical symmetry by representing the first order metric perturbation as a sum of tensor spherical-harmonic $lm$-modes in the frequency domain. Then, the construction of an $lm$-mode of the source can be treated as a Clebsch-Gordon problem in order to express a product of sums over modes as a sum over products of modes \cite{Spiers:2023mor}. The nonsmoothness of the source leads to strong nonlinear $lm$-mode coupling and poor convergence of the sum over modes. This can be made tolerable in the Schwarzschild case by using a puncture to isolate the most singular behavior and leave behind a more rapidly convergent residual sum \cite{Miller_2016}.

The reduced symmetry of Kerr spacetime makes the whole problem much more challenging. There are two new sources of difficulty, both of which make a full separation-of-variables approach much less practical:
\begin{enumerate}
    \item There is no known spheroidal equivalent of a tensor-spherical harmonic basis in which the metric perturbation equations separate. The direct metric perturbation problem can be formulated as a system of equations in which all $l$-modes are coupled together, which is an obstacle that would affect the field equations at both first and second order;
    \item The poor convergence arising from the nonsmoothness of the second-order source is exacerbated by the increased mode coupling. Although it is possible to formulate the source construction as a Clebsch-Gordon mode-coupling problem (see, for example, \ccite{Spiers_2024} for a demonstration in the case of the Teukolsky equation), this relies on a projection onto spin-weighted \emph{spherical} harmonics, which in turn introduces significant additional (and undesirable) coupling between modes.
\end{enumerate}
Work is underway to address some of these $lm$-mode challenges, for example addressing the first issue by formulating the problem of computing the metric perturbation in terms of reconstruction from spin-weighted scalar fields that satisfy separable Teukolsky equations~\cite{Green_2020,Toomani_2021,Bourg_2024,Dolan_2022,Dolan_2024,Wardell_2024}. However, none of these solutions addresses the problems inherent in constructing a second-order source for Kerr perturbations.

In this work we take a different approach and avoid $l$-modes altogether, instead pursuing a decomposition into $m$-modes only. There are two significant advantages to this approach: (i) We can adapt our discretization of the $(r,\theta)$ grid to the structure of the perturbation, which is approximately isotropic around the particle. This is in contrast to $l$-modes that are inherently adapted to the background Kerr spacetime and give a poorly converging representation of the perturbation; (ii) The construction of the source can be done pointwise in the $(r,\theta)$ grid, with coupling over $m$-modes only. It is still necessary to use a puncture to isolate the most singular behavior and leave a sufficiently rapidly converging residual sum, but the construction of the source is simpler: the $m$-mode approach involves a purely Fourier basis and it is easy (compared to $lm$-modes) to express products of Fourier series as a single Fourier series.

This work is part of a wider effort to develop a framework for using $m$-modes to solve the perturbation equations in Kerr spacetime.
The earliest $m$-mode studies were performed in the time domain~\cite{Barack_2007a,Barack_2007b,Dolan_2011a,Dolan_2011b,Dolan_2013,Thornburg_2017}; however,
with multiscale methods emerging as the most promising approach to solving the binary inspiral problem within black hole perturbation theory, focus has shifted toward frequency domain approaches \cite{Miller:2020bft,Pound_2021,Miller:2023ers,Mathews:2025nyb,Lewis:2025ydo}. In the context of $m$-modes, these involve separating the $t$ and $\varphi$ variables so that each $m$-mode of the field satisfies a linear elliptic partial differential equation (PDE) with $r$ and $\theta$ derivatives. Existing work has successfully developed: a framework for performing frequency-domain scalar self-force calculations using $m$-modes \cite{Osburn_2022}; hyperboloidal slicing and coordinate systems adapted to the particle \cite{Macedo_2024}; efficient methods for constructing puncture fields \cite{Bourg:2024cgh} and isolating the singular behavior of the second-order source \cite{Bourg_Capra_2025}; and tools for constructing the second order source through $m$-mode coupling \cite{Dyson_Capra_2025}. In this work we focus on another core part of the problem, leveraging computational tools and methods from the adjacent field of numerical relativity (NR), many of which we are introducing to perturbative Kerr calculations for the first time, to make these calculations computationally tractable (there is also time domain work involving overlap between self-force and numerical relativity~\cite{Vega_2009,Dhesi_2021,Wittek_2023,Wittek_2024a,Wittek_2024b}).

Specifically, we adapt computational methods that were originally developed to solve the elliptic Einstein constraint equations for the initial slice of spacetime at the start of every NR simulation of merging binary black holes. To solve these initial data problems, the NR community has developed sophisticated methods to solve elliptic PDEs on supercomputers~\cite{Vu_2022,Lorene,Ansorg:2004ds,Pfeiffer2003-mt,Papenfort:2021hod,Rashti:2021ihv,Uryu2012-bo,Assumpcao2021-es}. In particular, the ``puncture'' initial data formalism~\cite{Brandt:1997tf} presents similar computational challenges to the self-force problem, in that both use punctures to handle nonsmooth points in the domain.
Our implementation adapts the elliptic solver in the open-source numerical relativity platform \spectre{}~\cite{spectrecode,Vu_2022} to solve $m$-mode self-force problems.
The code uses discontinuous Galerkin (DG) methods with adaptive mesh refinement (AMR) and an iterative Krylov-type linear solver with multigrid-Schwarz preconditioning. These methods were shown to generalize to different elliptic problems in \ccite{Vu_2022}, including puncture initial data with arbitrarily placed punctures in \ccite{Vu:2024}. We extend the solver to handle complex-valued fields, self-force regularization through the effective source method~\cite{Barack_2007a,Barack_2007b,Dolan_2011a,Dolan_2011b,Dolan_2013,Thornburg_2017,Detweiler_2003,Vega_2008,Vega_2009,Vega_2011,Wardell_2012}, and preconditioning of the resulting complex-valued Helmholtz-type problems. We adopt a similar rectangular $(r,\theta)$ domain arrangement as the less accurate finite-difference calculation by Osburn and Nishimura~\cite{Osburn_2022}, but achieve exponential convergence due to our use of spectral DG methods and $hp$ AMR. Hyperboloidal slicing is an important feature of our implementation that avoids asymptotic oscillations (see~\cite{Macedo_2024} for an independent approach) and also enables horizon penetrating coordinates. Through these NR-based enhancements, we are able to calculate the conservative scalar self-force for both prograde and retrograde motion at the innermost stable circular orbit (ISCO) at spins up to $|a|/M=0.998$ (Thorne limit) in a few seconds per $m$-mode on a laptop computer. We sum over 20 $m$-modes, which are independent and hence parallelizable. To develop techniques in this article, we restrict focus to the more approachable scenario involving a scalar field generated by a scalar charge in a circular equatorial orbit around a Kerr black hole, but our method is constructed such that it can be generalized to gravitational self-force calculations and more generic orbits in the future.

This paper is organized as follows. \Cref{sec:formulation} develops the elliptic $m$-mode formulation of the scalar self-force equations. \Cref{sec:regularization} details the effective-source regularization method. \Cref{sec:solver} details the discretization of the equations and the iterative elliptic solver. \Cref{sec:results} presents our results. We conclude in \cref{sec:conclusion}.

\section{Scalar fields in Kerr spacetime}\label{sec:formulation}

\subsection{Circular equatorial geodesic motion}

We consider a scalar field sourced by a scalar point charge following a circular equatorial geodesic around a Kerr black hole with metric $g_{\mu\nu}$ given by
\begin{align}
ds^2 &= g_{\mu\nu} \, dx^\mu dx^\nu \notag
\\&= -\left(1-\frac{2Mr}{\Sigma}\right)dt^2-\frac{4Ma \, r\sin^2{\theta}}{\Sigma}d\phi \, dt + \frac{\Sigma}{\Delta} dr^2 \notag
\\&\;\;\;\; +\Sigma \, d\theta^2+\left( r^2+a^2+\frac{2Mr \, a^2\sin^2{\theta}}{\Sigma} \right) \sin^2{\theta} \, d\phi^2 \, ,
\end{align}
where $\Sigma \equiv r^2 + a^2 \cos^2{\theta}$ and $\Delta \equiv r^2-2Mr+a^2 = (r-r_+)(r-r_-)$, and where we have introduced Boyer-Lindquist coordinates $x^\alpha = (t, r, \theta, \phi)$ and the radii of the outer ($+$) and inner ($-$) horizons $r_\pm = M \pm \sqrt{M^2-a^2}$.

The position of the scalar charge is $x_p^\alpha$ with corresponding four-velocity
\begin{align}
u^\alpha = \frac{dx_p^\alpha}{d\tau} \, .
\end{align}
Equatorial motion involves two constants of motion associated with two Killing vectors $\xi_{(t)}^\alpha = \delta^\alpha_t$ and $\xi_{(\phi)}^\alpha = \delta^\alpha_\phi$ (with the Kronecker delta)
\begin{align}
\mathcal{E} &= -g_{\mu\nu} u^\mu \xi_{(t)}^\nu
\\&= \left(1-\frac{2M}{r_p}\right) u^t + \frac{2Ma}{r_p} u^\phi \, , \notag
\\
\mathcal{L} &= g_{\mu\nu} u^\mu \xi_{(\phi)}^\nu
\\&= \left( r_p^2+a^2+\frac{2Ma^2}{r_p} \right) u^\phi - \frac{2Ma}{r_p} u^t \, , \notag
\end{align}
where $\mathcal{E}$ is the specific energy, $\mathcal{L}$ is the specific angular momentum, and we have imposed $\theta_p = \pi/2$ (equatorial motion). The radial equation describing $u^r$ follows from $g_{\mu\nu} u^\mu u^\nu = -1$ with substitutions to eliminate $u^t$ and $u^\phi$ in favor of $\mathcal{E}$ and $\mathcal{L}$:
\begin{align}
& \frac{1-\mathcal{E}^2}{2} = \frac{1}{2}\left( u^r \right)^2 + V_\text{eff}(r_p) \, ,
\\
& V_\text{eff}(r_p) \equiv -\frac{M}{r_p} + \frac{\mathcal{L}^2+a^2(1-\mathcal{E}^2)}{2r_p^2} - \frac{M(\mathcal{L}-a\mathcal{E})^2}{r_p^3} \, .
\end{align}
For the circular case we pursue here, orbits can be uniquely parametrized by the constant orbital radius $r_p = r_0$, with $\mathcal{E}$ and $\mathcal{L}$ then determined in terms of $r_0$ through two conditions: $u^r=0$ and $\frac{d V_\text{eff}}{d r_p} = 0$. These circular orbit values of $\mathcal{E}$ and $\mathcal{L}$ further lead to corresponding constant values for $u^t$ and $u^\phi$:
\begin{align}
& u^t = \frac{a + \sqrt{r_0^3/M}}{\sqrt{r_0^3/M-3r_0^2+2a\sqrt{r_0^3/M}}} \, ,
\\
& u^\phi = \frac{1}{\sqrt{r_0^3/M-3r_0^2+2a\sqrt{r_0^3/M}}} \, ,
\end{align}
from which we determine the orbital angular frequency
\begin{align}
\Omega = \frac{d\phi_p}{dt} = \frac{u^\phi}{u^t} =  \frac{1}{a+\sqrt{r_0^3/M}} \, .
\end{align}
Since this is a constant, we can immediately integrate to get the orbital phase,
\begin{equation}
    \phi_p = \Omega\, t \, .
\end{equation}
According to this analysis, the sign of $a$ controls whether the motion is prograde or retrograde. There is an innermost stable circular orbit (ISCO) at the inflection point of $V_\text{eff}$. The orbital radius of this inflection point is called $r_\text{ISCO}$, and it is found from $\frac{d^2V_\text{eff}}{dr_p^2}=0$, which can be simplified to take the form
\begin{align}
0 = 1 - \frac{6M}{r_\text{ISCO}} + 8a\sqrt{\frac{M}{r_\text{ISCO}^3}} - \frac{3a^2}{r_\text{ISCO}^2} \, .
\end{align}

\subsection{Klein-Gordon equation in terms of $m$-modes}

The scalar charge produces a scalar field, $\Phi$, which we take to satisfy the massless Klein-Gordon equation
\begin{align}
\Box \Phi = -4\pi \rho \, ,\label{eq:KG}
\end{align}
where $\rho$ is the scalar charge density, $\Box\equiv g^{\mu\nu}\nabla_\mu\nabla_\nu$ and $\nabla_\alpha$ is the covariant derivative. The scalar charge density is that of a point particle with scalar charge $q$
\begin{align}
\label{eq:pp-source}
\rho = \frac{q}{u^t\sqrt{-g}} \, \delta(r-r_0) \, \delta(\theta-\pi/2) \, \delta(\phi-\Omega \, t),
\end{align}
where $g\equiv-\Sigma^2\sin^2{\theta}$ is the Kerr metric determinant.

The frequency domain $m$-mode strategy involves separating the $t$ and $\phi$ variables. We use Kerr's original azimuthal coordinate $\varphi$ to improve behavior near the horizon (following~\cite{Dolan_2011b})
\begin{align}
&\varphi(\phi,r) = \phi+\frac{a}{r_+-r_-}\ln{\left(\frac{r-r_+}{r-r_-}\right)} \, .
\end{align}
We also factor out $1/r$ and $\text{sin}^{|m|}\theta$ (following~\cite{Macedo_2024}) so that each mode approaches a nonzero constant amplitude as $r\to\infty$ and toward the poles, respectively, and we use the fact that for circular orbits we only need to consider frequencies $\omega = m \Omega$, so that our mode decomposition is given by:
\begin{align}
\label{eq:mode_sum}
&\Phi(t,r,\theta,\varphi) = \sum_{m={-\infty}}^\infty \tilde{\Psi}_m(r,\theta) \,\frac{\text{sin}^{|m|}\theta}{r} \, e^{im(\varphi-\Omega t)} \, ,
\end{align} 
where we have introduced the tilde in $\tilde{\Psi}_m$ to indicate mode functions that are based on $t$-slicing. The orthogonality of our $m$-mode basis provides the inverse relation
\begin{align}
\label{eq:modeIntegral}
\tilde{\Psi}_m = \frac{r\,e^{im\Omega t}}{2\pi\, \text{sin}^{|m|}\theta}\int_{-\pi}^{\pi} \Phi\,e^{-im\varphi} d\varphi \, .
\end{align}
We find it convenient to introduce a new angular variable (following~\cite{Macedo_2024})
\begin{align}
z=\cos{\theta} \, ,
\end{align}
to describe the polar domain and the functional dependence of $\tilde{\Psi}_m$.
Substituting Eq.~\eqref{eq:mode_sum} into Eq.~\eqref{eq:KG} reveals the elliptic PDE that governs each $\tilde{\Psi}_m(r,z)$:
\begin{align}\label{eq:elliptic_eqn}
\tilde{\triangle}_m \tilde{\Psi}_m = S_m \, ,
\end{align}
where
\begin{align}
&\tilde{\triangle}_m = -\frac{r^2+a^2}{\Delta}\frac{\partial^2}{\partial r_*^2} + \left( \frac{2a^2}{r(r^2+a^2)}-\frac{2ima}{\Delta} \right) \frac{\partial}{\partial r_*} \notag
\\& - \frac{1}{r^2+a^2} \bigg( \frac{\partial}{\partial z} (1-z^2) \frac{\partial}{\partial z} - 2mz\frac{\partial}{\partial z} - m(m+1) \notag
\\&\qquad\qquad + \frac{m^2\Omega\big(2Mr(\Omega(r^2+a^2)-2a)+\Omega\Delta\Sigma\big)}{\Delta} \notag
\\&\qquad\qquad\qquad\qquad +\frac{2(a^2-imar-Mr)}{r^2} \bigg) \, , \label{eq:triangle}
\\
&S_m = \frac{2r\Sigma\,e^{im\Omega t}}{(r^2+a^2)\,\text{sin}^{|m|}\theta}\int_{-\pi}^{\pi} \rho\,e^{-im\varphi} d\varphi \, ,
\end{align}
and $r_*$ is the tortoise coordinate
\begin{align}
r_* = r+\frac{2M}{r_+-r_-}\bigg[ r_+\ln{\Big( \frac{r-r_+}{2M} \Big)} - r_-\ln{\Big( \frac{r-r_-}{2M} \Big)} \bigg] \, ,
\end{align}
which is defined such that
\begin{align}
\frac{dr_*}{dr} = \frac{r^2+a^2}{\Delta} \, .
\end{align}
Notice that Eq.~\eqref{eq:elliptic_eqn} is an elliptic PDE because the coefficients of $\partial_{r_*}^2$ and $\partial_{z}^2$ always have the same sign.

In the following subsections we make some modifications to put this elliptic PDE into the form used in our numerical solver. Some of our choices in the definition of $\tilde{\triangle}_m$ have been made in anticipation of this. In particular, we note that the null behavior of $t\pm r_*$ makes the tortoise coordinate helpful in our upcoming discussion of hyperboloidal slicing, and this is why we use $\partial_{r_*}$ in Eq.~\eqref{eq:elliptic_eqn} despite the fact that our numerical discretization is based on $r$ rather than $r_*$. Similarly, we have chosen the overall scale factor applied to Eq.~\eqref{eq:elliptic_eqn} for future convenience in a way that builds toward our final form of the field equation.

\subsubsection{Hyperboloidal and \vtu{} slicing}

The formulation to this point, based on $t$-slicing and $\tilde{\Psi}_m$, is nearly equivalent to that of Osburn and Nishimura~\cite{Osburn_2022} (other than using $z$ rather than $\theta$). We attempted an associated numerical implementation using our iterative discretized elliptic PDE solver (see \cref{sec:solver}), but that early version of our code required a prohibitively large number of iterations to converge due to asymptotic oscillations associated with $t$-slicing\footnote{Osburn and Nishimura~\cite{Osburn_2022} did not encounter this issue because they applied a direct (rather than iterative) matrix solver to the linear algebra representation of their discretized field equations.}. This asymptotic wavelike behavior of $\tilde{\Psi}_m$  is a well-known feature of radiation problems. We eliminate these oscillations near the horizon and infinity by implementing hyperboloidal slicing, which is based on a radially dependent transformation of the $t$ coordinate:
\begin{align}
 s = t - h(r_*) \, .
\end{align}
The height function, $h(r_*)$, is chosen so that the new coordinate, $s$, is asymptotically null toward both the horizon and infinity.
The hyperboloidal condition can be conveniently expressed in terms of the boost function, $H$:
\begin{align}
&H(r_*) \equiv \frac{dh}{dr_*} \, ,
\\
&H(\pm \infty)=\pm 1 \, .
\end{align}
In the source region we enforce $s=t$ by having $H=0$, so $H$ will transition from -1 to 0 to +1 over the entire radial domain. For mode functions that are hyperboloidally sliced we omit the ``tilde" and introduce $\Psi_m(r,z)$ defined by the hyperboloidally-sliced mode decomposition
\begin{align}
\label{eq:hyperboloidal_mode_sum}
&\Phi(s,r,z,\varphi) = \sum_{m={-\infty}}^\infty \Psi_m(r,z) \,\frac{\text{sin}^{|m|}\theta}{r} \, e^{im(\varphi-\Omega s)} \, .
\end{align}
A comparison of Eqs.~\eqref{eq:mode_sum} and~\eqref{eq:hyperboloidal_mode_sum} provides the relationship between $\Psi_m$ and $\tilde{\Psi}_m$:
\begin{equation}
\label{eq:toTilde}
 \Psi_m (r,z)= \tilde{\Psi}_m (r,z) \, e^{-i m \Omega \, h(r_*)} \, .
\end{equation}

To enable calculations of $\Psi_m$, we transform $\tilde{\triangle}_m$ to account for the change to hyperboloidal coordinates; the multivariable chain rule provides this transformation. Although $\partial_t$ simply becomes $\partial_s$ under the coordinate transformation, transforming $\partial_{r_*}$ at fixed $t$ to $\partial_{r_*}$ at fixed $s$ introduces a term related to the boost function
\begin{align}
 \left[\frac{\partial}{\partial r_*}\right]_t &= \left[\frac{\partial}{\partial r_*}\right]_s + \left[\frac{\partial s}{\partial r_*}\right]_t \left[\frac{\partial}{\partial s}\right]_{r_*} \notag \\
&= \left[\frac{\partial}{\partial r_*}\right]_s -H \left[\frac{\partial}{\partial s}\right]_{r_*},
\end{align}
where the subscript indicates which coordinate is being held constant during the derivative.

At this stage, where each partial derivative holds all other newly transformed coordinates constant, it is appropriate to omit the square brackets and coordinate subscripts and think of these as replacements for the partial derivatives that appear in the field equations
\begin{align}
& \frac{\partial}{\partial r_*} \rightarrow \frac{\partial}{\partial r_*} - H \frac{\partial}{\partial s} \, ,
\\
&\frac{\partial^2}{\partial r_*^2} \rightarrow \frac{\partial^2}{\partial r_*^2} -2 H\frac{\partial^2}{\partial r_* \partial s}  -H' \frac{\partial}{\partial s} + H^2 \frac{\partial^2}{\partial s^2} \, ,
\end{align}
which, when re-entering the frequency domain (i.e.~replacing $s$ derivatives with $-im\Omega$ factors) becomes
\begin{align}
& \frac{\partial}{\partial r_*} \rightarrow \frac{\partial}{\partial r_*} + i m\Omega H \label{eq:rstarderiv-transformation}\, ,
\\
&\frac{\partial^2}{\partial r_*^2} \rightarrow \frac{\partial^2}{\partial r_*^2} +2im\Omega H\frac{\partial}{\partial r_*}  +im\Omega H' -m^2\Omega^2 H^2 .
\end{align}
Note that $H'$ represents a derivative with respect to $r_*$. Inserting these transformations into Eq.~\eqref{eq:triangle} produces a new differential operator we refer to as $\triangle_m$ (where the ``tilde" is now omitted)
\begin{align}
& \triangle_m \Psi_m = S_m \, . \label{eq:triangle_hyp}
\end{align}

There is much freedom in choosing the way $H$ transitions from $-1$ to $0$ to $+1$. Both smooth transitions and certain nonsmooth transitions are valid choices. In this work we use a specific nonsmooth choice of piecewise-constant boost function referred to as ``\vtu{}" slicing:
\begin{align}
H(r_*) = \Bigg\{ \begin{array}{rll}
-1, &\quad r_* \le r^v_* &\quad \text{($v$ domain)} \\
0, &\quad r^v_* \le r_* \le r^u_* &\quad \text{($t$ domain)} \\
+1, &\quad r_* \ge r^u_* &\quad \text{($u$ domain)}
\end{array} ,
\end{align}
where $r^v_*$ and $r^u_*$ are constants that satisfy $r^v_* < r_{*0} < r^u_*$. Note that we sample the interfaces between the domains from both sides, so our field is essentially double valued there---this is a deliberate choice so that our DG scheme can apply junction conditions at these interfaces. The definition in terms of a boost function only determines the height function $h(r_*)$ up to domain-dependent constants of integration; we choose these such that the height function is continuous across domain interfaces:
\begin{align}
h(r_*) = \Bigg\{ \begin{array}{rll}
-(r_*-r_*^v), &\quad r_* \le r^v_* &\quad \text{($v$ domain)} \\
0, &\quad r^v_* \le r_* \le r^u_* &\quad \text{($t$ domain)} \\
+(r_* - r_*^u), &\quad r_* \ge r^u_* &\quad \text{($u$ domain)}
\end{array} .
\end{align}
One of the advantages of $vtu$ slicing is simplicity: there are only two free parameters ($r^v_*$ and $r^u_*$) that need to be tuned to maximize the accuracy of the solution. This is in contrast to smooth transition functions, where the choice of transition function can introduce many more parameters (e.g. type of function, width of the transition region, steepness of the transition function) that must be fine-tuned to the problem at hand. Smooth transition functions also tend to introduce additional structure in the solution; this additional structure must be resolved in the numerical solution and thus can have a negative effect on accuracy.

The jump in $H$ produces a jump in the radial derivative of $\Psi_m$ at the transition from $v$ to $t$, and again at the transition from $t$ to $u$; we account for these jumps by imposing appropriate junction conditions at the interfaces between domains.
First, note that the fundamental behavior of $\Phi$ is not directly affected by slicing; assuming events are correctly matched across slicings, the values of $\Phi$ at each event are the same. Since $\Phi$ and $t$ are both smooth functions away from the particle worldline, this then implies that $\tilde\Psi_m$ is also a smooth function. We next consider how $t$-sliced mode functions $\tilde\Psi_m$ are related to $s$-sliced mode functions $\Psi_m$ through Eq.~\eqref{eq:toTilde}.
Since $h(r_*)$ is continuous across domain interfaces, we have that $\Psi_m$ is also continuous across interfaces.
The same is not true for the radial derivative of $\Psi_m$; since $\tilde{\Psi}_m$ is a smooth function, Eq.~\eqref{eq:rstarderiv-transformation} implies that $\partial_{r_*} \Psi_m$ has a jump whenever $H(r_*)$ does. In particular, this implies that the jump in the derivative is
\begin{align}\label{eq:jump_vt}
\frac{\partial\Psi_m}{\partial r_*}\bigg|_{r_*=(r^{v}_*)^+} - \frac{\partial\Psi_m}{\partial r_*}\bigg|_{r_*=(r^{v}_*)^-} = -im\Omega\Psi_m \big|_{r_*=r^v_*}\, ,
\end{align}
at the transition from $v$ to $t$ and that it is
\begin{align}\label{eq:jump_tu}
\frac{\partial\Psi_m}{\partial r_*}\bigg|_{r_*=(r^{u}_*)^+} - \frac{\partial\Psi_m}{\partial r_*}\bigg|_{r_*=(r^{u}_*)^-} = -im\Omega\Psi_m \big|_{r_*=r^u_*}\, .
\end{align}
at the transition from $t$ to $u$ .
Our DG numerical method is well suited to account for such jumps.%

Finally, we note that our choice of \vtu{} slicing causes the $H'$ terms in the transformed field equation to vanish everywhere except at the transition radii, and that these points are handled by junction conditions.

\subsubsection{Horizon-penetrating coordinates and compactification}

The elimination of asymptotic oscillations via hyperboloidal slicing enables compactification of the radial coordinate. This brings several advantages including more efficient discretizations (by focusing resolution where it is most needed) and a simplified boundary treatment.
We also found that compactifying toward the horizon remedies an issue encountered in the case of gravitational perturbations, which will be described in future work.

Because the Boyer-Lindquist $r$ coordinate is inherently compact near the horizon, our compactification begins by discretizing based on $r$ and expressing the differential operator in terms of $r$ derivatives,
\begin{align}
    \frac{\partial}{\partial r_*} = \frac{\Delta}{r^2 + a^2} \frac{\partial}{\partial r}
    \, \text{.}
\end{align}
This transformation places the horizon at the finite radius $r=r_+$ rather than at $r_* \rightarrow -\infty$. Together with hyperboloidal slicing, the result is a horizon-penetrating ingoing Eddington-Finkelstein coordinate system near the horizon ($v$ domain), simple Boyer-Lindquist radial and time coordinates near the puncture ($t$ domain), and outgoing Eddington-Finkelstein coordinates in the wave zone ($u$ domain).

We also partially compactify at large radius by introducing a $\sim 1/r$ coordinate map in the $u$ domain; further details on this are given in \cref{sec:discretisation}.

\subsubsection{Equations in first-order flux form}

Our final form of the field equation enables the use of DG methods by expressing the equations in first-order form with a divergence producing the principal part (see~\cref{sec:solver}). To manipulate the field equations into this form, it is convenient to introduce the 2D coordinates $q^i \equiv (r, z)$ and to define the 2D flux vector field $\mathcal{F}^i_m$,
\begin{align}\label{eq:flux}
\mathcal{F}^i_m \equiv \left(\frac{\Delta}{r^2+a^2}\frac{\partial \Psi_m}{\partial r}\;,\,\frac{1-z^2}{r^2+a^2}\frac{\partial \Psi_m}{\partial z} \right) \, .
\end{align}
Now the PDE can be expressed in the following first-order form involving a divergence:
\begin{align} \label{eq:first-order}
\triangle_m\Psi_m = -\partial_i \mathcal{F}^i_m + \beta_m \Psi_m + \gamma^i_m \partial_i \Psi_m \, ,
\end{align}
where $\partial_i \equiv \frac{\partial}{\partial q^i}$, the 2D divergence $-\partial_i \mathcal{F}^i$ produces the principal part of $\triangle_m\Psi_m$, and the coefficients $\beta_m$ and $\gamma^i_m$ are functions of $r$ and $z$. The functional forms of $\beta_m$ and $\gamma^i_m$
are given by
\begin{align}
    \beta_m =& \frac{1}{r^2 + a^2}\bigg[m(m+1) + \frac{2M}{r}\Big(1-\frac{a^2}{Mr}\Big) 
\\&\qquad\qquad\qquad\qquad\qquad + \frac{2iam}{r} (1 + a\Omega H) \bigg] \notag
\\&+ \frac{r^2 + a^2}{\Delta} \bigg[m^2\Omega^2 \Big(H^2 - 1 + \frac{a^2 \Delta (1-z^2)}{(r^2 + a^2)^2}\Big) \notag
\\&\qquad\qquad + \frac{2am^2\Omega}{r^2+a^2}\Big(\frac{2Mr}{r^2+a^2} + H\Big) - im\Omega H' \bigg] \, , \notag \\
    \gamma^r_m = &2\left[\frac{a^2\Delta}{r (r^2+a^2)^2} - \frac{ima}{r^2+a^2} - im\Omega H\right] \, , \\
    \gamma^z_m = &\frac{2mz}{r^2 + a^2}
    \text{.}
\end{align}

Notice that at the boundaries corresponding to the horizon ($r=r_+$) and the poles ($z^2=1$) the flux component normal to that boundary (either $\mathcal{F}^r_m$ or $\mathcal{F}^z_m$, respectively) will vanish.
Also, although it may appear that $\beta_m$ behaves poorly approaching the horizon, the $1/\Delta$ terms cancel when $H=-1$ such that $\beta_m$ has a finite value at the horizon.
These asymptotic properties at the boundaries affect how we approach boundary conditions in Sec.~\ref{sec:bcs}, and it is favorable features such as these that motivated certain earlier choices including the scale factor for $\triangle_m$.

Finally, we may impose equatorial symmetry in our equations since we consider only equatorial orbits in this work. This is optional in our method and we have not found that it leads to significant improvements in the convergence of the solver or its time to solution. To impose equatorial symmetry we simply formulate the equations in the coordinate $z^2$ rather than $z$, with corresponding modifications to the coefficients:
\begin{align}
\mathcal{F}_m^{z^2} &= 4z^2\,\mathcal{F}_m^z \, , \\
\gamma_m^{z^2} &= 2z\,\gamma_m^z + 2\frac{1-z^2}{r^2 + a^2} \,
\text{.}
\end{align}

\subsubsection{Boundary conditions}
\label{sec:bcs}

Our formulation of the field equation avoids coordinate singularities at the horizon and reduces %
boundary conditions to simple regularity conditions. The vanishing of $\mathcal{F}^r_m$ at the horizon means that we do not have to explicitly impose boundary conditions there. Similarly, factoring out the $\text{sin}^{|m|}\theta$ behavior in \cref{eq:mode_sum} causes $\mathcal{F}^z_m$ to vanish at the poles such that the $z$ boundaries do not require explicit boundary treatment either (it can be shown that this is equivalent to guaranteeing $\Phi$ is continuous and sufficiently differentiable at the poles, see~\cite{Osburn_2022}).
These regularity conditions are particularly suitable for spectral numerical methods, because they expand fields in a polynomial series that inherently selects regular solutions.

In our implementation (see \cref{sec:solver}) we place the outer radial boundary at a large but finite radius rather than null infinity. To do so, we impose a second-order Bayliss-Turkel boundary condition (following~\cite{Osburn_2022}),
\begin{align}
    \partial^2_{r_*} \tilde\Psi_m - 2im\Omega\,\partial_{r_*} \tilde\Psi_m - m^2\Omega^2 \tilde\Psi_m = 0 \,
    \text{.}
\end{align}
Notice that this is given in terms of the $t$-sliced $\tilde\Psi_m$, and transforming to $u$-slicing provides the equivalent condition
\begin{align}\label{eq:bayliss_turkel_bc}
    \partial^2_{r_*} \Psi_m = 0 \,.
\end{align}
The numerical implementation of these boundary conditions is described in \cref{sec:bc_impl}.

\section{Self-force regularization}\label{sec:regularization}

So far we have focused our attention on the differential operator $\triangle_m$. We now shift focus to the source of the differential equation.

\subsection{Puncture and residual fields}

The retarded solution to the Klein-Gordon equation %
with a point-particle source %
diverges as $\Phi \sim 1/\lambda$, where $\lambda$ is a formal order-counting parameter that counts powers of spatial distance from the particle’s worldline. The corresponding $m$-modes of the retarded field, $\Psi_m$, diverge logarithmically toward the worldline. This causes two issues: 1) the field is infinite within the domain where we seek numerical solutions and 2) calculating the self-force requires information at precisely the position where the scalar field is infinite. Both issues are resolved by introducing a \textit{puncture} field $\Phi^{\cal{P}}$ that locally captures the singular behavior. Subtracting the puncture field from the retarded field we get a \textit{residual} field:
\begin{align}
\label{eq:singular-regular}
\Phi^{\cal{R}} = \Phi - \Phi^{\cal{P}}.
\end{align}
For an appropriate choice of puncture, each of the three fields possesses valuable features: $\Phi$ obeys well-defined retarded boundary conditions, $\Phi^{\cal{P}}$ isolates the singular behavior near the worldline,
and $\Phi^{\cal{R}}$ is finite on the worldline and quantifies the self-force, 
\begin{align}
F^\text{self}_\alpha = q\,\nabla_\alpha \Phi^{\cal{R}} \big|_{x^\mu = x^\mu_p}.
\end{align}
Substituting Eq.~\eqref{eq:singular-regular} into the Klein-Gordon equation, Eq.~\eqref{eq:KG}, we obtain an equation for the residual field,
\begin{equation}
    \Box \Phi^{\cal R} = -4\pi \rho - \Box \Phi^{\cal P} \coloneqq S^{\rm eff}.
\end{equation}
This is another Klein-Gordon equation, but with a more well-behaved \emph{effective source} \cite{Barack_2007b,Vega_2008,Vega_2009,Vega_2011,Wardell_2012}:
by construction, the puncture field is defined so that the Dirac delta functions on the worldline that appear in $\rho$ are exactly canceled, leaving an effective source that is finite everywhere. 

In this work, we adopt a puncture field that is a local, power-series approximation to the Detweiler-Whiting~\cite{Detweiler_2003} singular field in the vicinity of the worldline. Our puncture is the same as the one described in \ccite{Thornburg_2017} with some tweaks to the implementation to resolve numerical accuracy issues. More specifically, we adopt a fourth-order puncture (keeping four terms in the local series expansion, $\Phi^{\cal P}\sim 1/\lambda + \lambda^0 + \lambda^1 +\lambda^2$), we use the ``Q-R'' scheme from \ccite{Wardell_2012}, and we apply the methods of \ccites{Heffernan_2012,Heffernan_2014} to obtain a four-dimensional puncture field given by
\begin{align}
\label{eq:puncture}
\Phi^{\cal P}(x^\alpha; x^\alpha_p) = \frac{1}{R^7} \sum_{i,j,k} A_{ijk}(x_p^\alpha, u^\alpha) \Delta r^i \Delta \theta^j Q^k,
\end{align}
where $\Delta r \coloneqq r-r_0$, $\Delta \theta \coloneqq \theta - \pi/2$, $Q\coloneqq\sin (\Delta \varphi/2)$, and
\begin{equation}
\label{eq:s2}
    R^2 \coloneqq \varrho^2 + z_c^2\, Q^2
\end{equation}
with
\begin{subequations}
\begin{align}
    \varrho^2 &\coloneqq \frac{r_p^2}{r_p^2-2M r_p+a^2} \Delta r^2 + r_p^2 \Delta \theta^2, \\
    z_c^2 &\coloneqq 4 ({\cal L}^2 + r_p^2 + a^2 + \frac{2M a^2}{r_p}).
\end{align}
\end{subequations}
The sum here is over all $i$, $j$ and $k$ such that $6\le i+j+k \le 9$.
Note that the coefficients $A_{ijk}$ are functions of the particle position and four-velocity only, and that for our case of circular equatorial orbits we have that $A_{ijk}=0$ whenever either $j$ or $k$ is odd.

\subsection{Mode decomposition}
Next we want to calculate the $m$-mode punctures $\Psi_m^\mathcal{P}$ from $\Phi^\mathcal{P}$ according to the $\varphi$ integral in~\cref{eq:modeIntegral} (note that the same logic defines $\Psi_m^\mathcal{R}$ from $\Phi^\mathcal{R}$). Substituting Eq.~\eqref{eq:puncture} into Eq.~\eqref{eq:modeIntegral} and using Eq.~\eqref{eq:s2} to eliminate $Q$ in favor of $R$, we obtain integrals that are of the form
\begin{equation}
    \int_{-\pi}^\pi R^n e^{-i m \Delta\varphi} d\Delta\varphi
\end{equation}
with $n$ an odd integer.
In \ccite{Thornburg_2017} analytic expressions for these integrals were found in terms of complete elliptic integrals multiplied by $m$-dependent polynomials in $\varrho/z_c$. Unfortunately, that approach introduces undesirable numerical noise in the puncture far from the worldline: the polynomials are ill conditioned for sufficiently large $\Delta r$ and/or $\Delta \theta$, and this problem becomes particularly pronounced for large $m$, since the degree of the polynomial grows with increasing $m$.

To mitigate this problem we instead use Eq.~(115) of \ccite{Bourg:2024cgh} to express the integrals as
\begin{align}
&\int_{-\pi}^\pi R^n e^{-i m \Delta\varphi} d\Delta\varphi =
\\&\qquad\qquad\qquad \frac{2 \pi\,(-1)^m }{(1+n/2)_m} \left[\varrho \sqrt{\varrho^2 + z_c^2}\right]^{n/2} {}_3 P_{n/2}^m(z) \, ,  \notag
\end{align}
where $z \coloneqq (\varrho^2 + z_c^2/2)/(\varrho \sqrt{\varrho^2 + z_c^2})$, $(x)_m$ is the Pochhammer symbol applied to $x$ with $m$ factors, and ${}_3 P_n^m(z)$ is the associated Legendre function of type 3, which is defined for $z>1$.
This can be accurately evaluated far from the worldline without being impacted by the roundoff errors that affect direct polynomial evaluation.

We numerically evaluate the associated Legendre function using its expression in terms of a regularized hypergeometric function:
\begin{align}
    {}_3 P_{n}^m(z) &= \frac{\Gamma(n+m+1)}{\Gamma(n-m+1)}\left(\frac{z-1}{z+1}\right)^{m/2}\left(\frac{2}{z+1}\right)^{n+1} \nonumber \\
    &\qquad{}_2\mathbf{F}_1\left(n+m+1, n+1, m+1, \frac{z-1}{z+1}\right),
\end{align}
which is valid when both $n \ge 0$ and $m\ge 0$. Cases where $n<0$ are evaluated using the identity ${}_3 P_{n}^m(z) = {}_3 P_{-n-1}^m(z)$ and cases where $m<0$ are evaluated using the identity ${}_3 P_{n}^{-m}(z) =\frac{\Gamma(n-m+1)}{\Gamma(n+m+1)}{}_3 P_{n}^m(z)$. We then make use of the \texttt{gsl\_sf\_hyperg\_2F1\_renorm} function from the GNU Scientific Library \cite{GSL} to evaluate the hypergeometric function ${}_2\mathbf{F}_1$. This works reliably everywhere except close to the worldline, where $z$ becomes large, the argument of the hypergeometric function gets close to $1$, and the \texttt{gsl\_sf\_hyperg\_2F1\_renorm} function fails to produce an accurate value. We therefore switch to computing the hypergeometric function using a direct evaluation of Eqs.~(\href{https://dlmf.nist.gov/15.8.10}{15.8.10}) and (\href{https://dlmf.nist.gov/15.8.12}{15.8.12}) of \ccite{NIST:DLMF} whenever $\frac{z-1}{z+1} \ge 0.995$. The sums in these expressions converge rapidly and are truncated as soon as the contribution from a term is less than machine epsilon relative to the total value of the sum.

We require derivatives of the puncture for the effective source. The $r$ and $\theta$ derivatives are easily computed analytically using the fact that the dependence on $r$ and $\theta$ is only through $\Delta r$, $\Delta \theta$ and $\varrho$, along with the fact that $\nabla_\alpha \varrho^n = \frac{n}{2} \varrho^{n-2} \nabla_\alpha \varrho^2$ and
\begin{equation}
\dfrac{d\,{}_3P_n^m(z)}{dz} = \frac{(n-m+1)\, {}_3P_{n+1}^m(z)-(n+1) z\, {}_3P_n^m(z)}{z^2-1}.    
\end{equation}
The $r$ and $\theta$ derivatives are therefore analytic expressions similar to the puncture itself. In our $m$-mode scheme, $t$ and $\varphi$ derivatives are trivially given by multiplication by $-i m\Omega$ and $i m$, respectively.\footnote{In principle there is also time dependence in the coefficients $A_{ijk}$ through their dependence on the worldline, but for circular equatorial orbits this can be ignored since $u^\alpha$, $r_p$ and $\theta_p$ are constant in that case, and thus so are the coefficients.}

There is one final numerical issue that affects the evaluation of the effective source. The effective source consists of a sum of terms involving the second derivative of the puncture, each of which diverges as $1/\lambda^2$. However, by construction four orders of singularity are canceled when these terms are combined to produce the effective source. This cancellation introduces roundoff error that becomes significant for points very close to the worldline. We therefore switch to using a local power-series approximation (in which the singular terms can be canceled analytically) for the small number of points where roundoff error would be significant. Our power series is accurate through order $\lambda^2$ and is of the form
\begin{align}
    &
    S^{[0]} \lambda^0 + S^{[1]}_1 \Delta r \lambda^1
    + \Big[S^{[2]}_2\Delta r^2 + S^{[2]}_0 \varrho^2
    \nonumber\\&\quad
    + (S^{[2]}_{2L}\Delta r^2 + S^{[2]}_{0L}\varrho^2) \log \varrho
    \nonumber\\&\quad
    + S^{[2]}_{4}\frac{\Delta r^4}{\varrho^2}
    + S^{[2]}_{6}\frac{\Delta r^6}{\varrho^4}
    + S^{[2]}_{8}\frac{\Delta r^8}{\varrho^6}
    \nonumber\\&\quad
    + S^{[2]}_{10}\frac{\Delta r^{10}}{\varrho^8}
    + S^{[2]}_{12}\frac{\Delta r^{12}}{\varrho^{10}}
    \Big]\lambda^2+\mathcal{O}(\lambda^3).
\end{align}
The coefficients, given explicitly in the \codename{EffectiveSource} repository \cite{EffectiveSource}, are algebraic functions of $r_0$. Their $m$ dependence is as follows: $S^{[0]} \propto \frac{1}{(2m-1)(2m+1)}$ and $S^{[1]} \propto \frac{1}{(2m-1)(2m+1)}$, $S^{[2]}_0$ and $S^{[2]}_2$ are rational polynomials in $m$ and also have terms involving the harmonic number $H_{-m-1/2}$, and all other coefficients are independent of $m$.

\subsection{Worldtube effective source}

It may seem like $\Psi_m^\mathcal{R}$ would be our most favorable target for numerical calculations because of its good behavior near the particle; however, $\Psi_m^\mathcal{R}$ does not satisfy predictable boundary conditions. In contrast, $\Psi_m$ obeys well-defined boundary conditions as the retarded field. Because of this, we introduce a rectangular worldtube (in the 2D $r$ and $\theta$ sense) centered on the particle. We solve for $\Psi_m^\mathcal{R}$ in the ``interior'' region inside the worldtube, we solve for $\Psi_m$ in the ``exterior'' region outside the worldtube, and we apply discontinuous jumps from $\Psi_m$ to $\Psi_m^\mathcal{R}$ at the edges of the worldtube. Similar to the $vtu$ slicing jumps, these worldtube jumps from $\Psi_m$ to $\Psi_m^\mathcal{R}$ are straightforward to implement within our DG scheme (see \cref{sec:jump_cond}). This leads to the following global approach to the $m$-mode field equations
\begin{align}
\triangle_m\,\Psi_m^\mathcal{R} &= S_m^\text{eff} \qquad\qquad\;\; \text{(interior),}
\\
\triangle_m\,\Psi_m &= 0  \qquad\qquad\;\;\;\;\;\; \text{(exterior),}
\\\label{eq:jump_worldtube}
\Psi_m^\mathcal{R} &= \Psi_m - \Psi_m^\mathcal{P} \qquad \text{(across interface).}
\end{align}
\subsection{Smoothness of the residual field} \label{sec:smoothness}
Recalling that our puncture is only a local approximation to a given order, this leads to a residual field that is not perfectly smooth at the particle. In our case we are using a 4D puncture that is valid through order $\lambda^2$, so the corresponding residual field is nonsmooth at order $\lambda^3$ and the effective source is nonsmooth at order $\lambda$. In $m$-modes, the puncture captures nonsmooth behavior in the retarded field of the form $\Psi^{\cal P}_m \sim\log \lambda + \lambda (1+\log \lambda) + \lambda^2 (1+\log \lambda) + \lambda^3 (1+\log \lambda)$ so the residual field has nonsmoothness of the form $\lambda^4 (1+\log\lambda)$, and the effective source has a $\lambda^2 (1+\log\lambda)$ nonsmoothness. The nonsmoothness also affects the convergence of the sum over modes: when evaluating the residual field on the worldline the sum over $m$-modes converges polynomially as $m^{-4}$ \cite{Thornburg_2017}. This convergence can be accelerated using analytically derived or numerically fitted regularization parameters that capture the polynomial behavior \cite{Heffernan_2014}.

\section{Numerical methods}\label{sec:solver}

\begin{figure*}
    \centering
    \includegraphics[width=\textwidth,clip,trim=10cm 8.5cm 14cm 10cm]{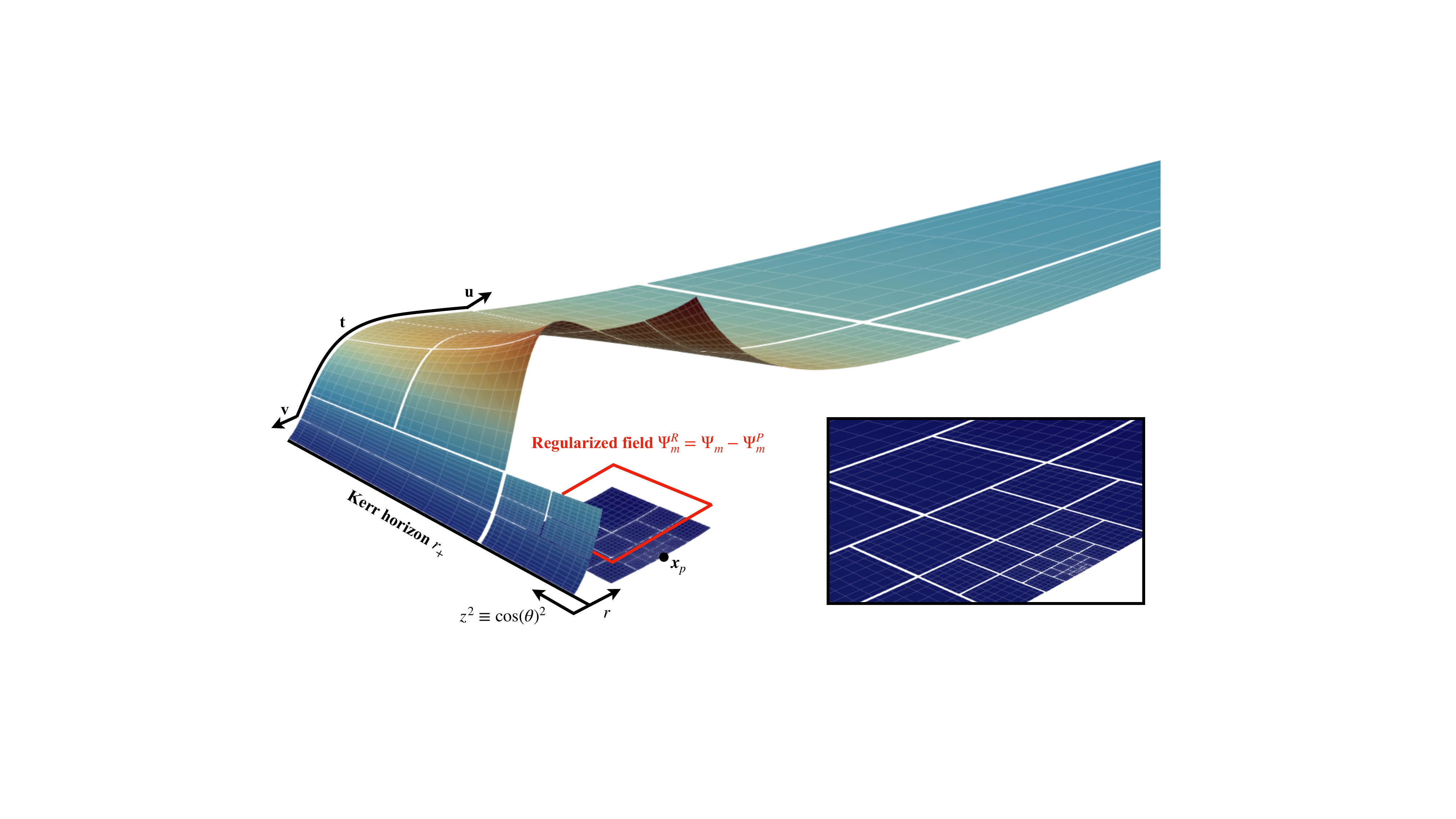}
    \caption{Computational domain for $a=0.5 M$, $r_0=10 M$, $m=1$ composed of discontinuous Galerkin elements. The domain extends from the Kerr horizon $r_+$ to the outer boundary at $R=10^4 M$, and from $z^2=0$ (equator) to $z^2=1$ (poles) with equatorial symmetry (optional). Plotted is the complex magnitude of the field $\Psi_m$. In the regularized region (red rectangle), the complex magnitude of the regularized field $\Psi^\mathcal{R}_m=\Psi_m-\Psi^\mathcal{P}_m$ is plotted instead. Null slicing is applied in the $v$ and $u$ regions with boost function $H=\pm 1$. The inset magnifies the region around the position of the puncture, $\bm{x}_p$.}
    \label{fig:domain}
\end{figure*}

Our formulation of the self-force problem results in a complex-valued linear elliptic PDE for each $m$-mode, $\triangle_m \Psi_m = S_m$. To solve the problem numerically, we have to choose a discretization method that turns the elliptic PDE into a matrix equation, $A\cdot u=b$, and a linear solver method to invert the matrix equation, schematically $u=A^{-1} \cdot b$, where the state vector $u$ represents all discrete degrees of freedom (typically field values of $\Psi_m$ on a grid). Both choices have complexity and depth. Fortunately, the field of numerical relativity, which is thematically adjacent to gravitational self-force theory and also in rapid development to gear up for the next generation of gravitational-wave astronomy, has a long history of solving elliptic PDEs to generate initial data for binary black hole simulations and substantial overlap in both expertise and personnel. Therefore, we propose a synergistic approach in which we adapt the elliptic solver from the open-source numerical relativity platform \spectre{}~\cite{spectrecode} to solve self-force problems. This approach provides us with a high-performance iterative elliptic solver based on discontinuous Galerkin (DG) methods. The DG discretization of elliptic equations is detailed in \ccites{Vu:2021,Vu:2024}, and the iterative elliptic solver is detailed in \ccite{Vu_2022}. Here, we give an overview of these methods and detail the advancements we made to the solver in order to solve self-force problems with the numerical relativity code.

\subsection{Discontinuous Galerkin discretization}
\label{sec:discretisation}

Discontinuous Galerkin methods are a class of spectral finite-element discretization methods for PDEs. They converge exponentially fast for smooth problems while also supporting a domain decomposition into many sparsely coupled elements suitable for parallelization. They are particularly well suited for our elliptic self-force calculations because discontinuities in our problem are static and known analytically and can often be placed at grid boundaries, retaining the exponential convergence (even at the locations of the puncture, regularized region, and slicing transitions). Having formulated the self-force problem in first-order flux divergence form, \cref{eq:first-order}, we can readily use the elliptic DG methods implemented in \spectre{}~\cite{Vu:2021,Vu:2024}.
The advancements we had to implement in \spectre{} to support elliptic self-force calculations, in addition to providing the coefficients and sources in \cref{eq:first-order}, were (1) generic support for complex-valued fields throughout the code, and (2) support for modifying DG fields and fluxes across element boundaries (\cref{sec:jump_cond} below). The DG methods are detailed in \ccites{Vu:2021,Vu:2024} and summarized in this section.

\subsubsection{Domain decomposition}

\Cref{fig:domain} gives an overview of our DG domain decomposition. We split the two-dimensional rectangular domain with coordinates $q^i=(r, z)$ (or $q^i=(r, z^2)$ for equatorial symmetry) in $5 \times 3$ irreducible \emph{blocks} (or $5 \times 2$ for equatorial symmetry) to define the different regions of our problem: in the radial direction we have a horizon-penetrating wave zone ($v$-slicing), a transition region ($t$-slicing), the worldtube region ($t$-slicing), another transition region ($t$-slicing), and an outer wave zone ($u$-slicing). In the angular direction we just have the worldtube and the polar boundary region(s). The locations of these block boundaries are chosen as input parameters and depend on the orbital configuration ($a$ and $r_0$) but remain fixed for different $m$. We choose the block boundaries such that the worldtube is large enough to avoid steep gradients in the solution near the puncture, and small enough to minimize computational expense from resolving small-scale features in the residual field.

Each block carries a coordinate map to ``logical'' coordinates $\xi^i=(\xi,\eta) \in [-1,1]^2$. This map is linear in the $v$ and $t$ regions in both radial and angular dimension, but it is radially proportional to $1/r$ in the $u$ region to resolve the large wave zone. Specifically, the radial coordinate map in the wave zone is
\begin{align}
    r(\xi) = \frac{2 r_u R}{r_u + R + (r_u - R)\xi}
    \text{,} \qquad \text{($u$ domain)} \label{eq:compaction}
\end{align}
where $r_u$ is the transition radius to the wave zone and $R=10^4 M$ is the outer radius. This map is apparent in \cref{fig:domain} in the $u$ region where the spacing between grid points increases radially in the wave zone.

\subsubsection{Mesh refinement}\label{sec:refinement}

We split blocks into many smaller \emph{elements}, $\Omega_k$, by repeatedly dividing the logical cube in half along either of the two dimensions $\xi$ and $\eta$ ($h$ refinement). Specifically, our $h$ refinement strategy is to split all elements containing the puncture in half in both dimensions in each refinement step. Since we place the particle at the center of the regularized worldtube, all four corner neighbors of the particle (two in case of equatorial symmetry) will split in half in each refinement step. When splitting elements in half, we retain a two-to-one balance at element interfaces, meaning that neighboring elements can differ by at most one $h$ refinement level in any dimension. In cases where the two-to-one balance would be violated, the offending elements are split in half as well to restore the two-to-one balance. The result of this $h$ refinement strategy after six refinement steps is shown in \cref{fig:domain}, with the region around the puncture magnified in the inset. Note that blocks neighboring the worldtube were split to retain the two-to-one balance.

Within each element, we expand fields in a polynomial basis ($p$ refinement). To this end, we define a grid of $N=N_\xi \times N_\eta$ Legendre-Gauss collocation points, $\bm{\xi}_p=(\xi_{p_1}, \eta_{p_2})$, in each element (visible as grid lines within each element in \cref{fig:domain}). On this grid, we expand fields in a Lagrange polynomial basis
\begin{align}\label{eq:spectral_sum}
    u(\bm{q}) = \sum_p^N u_p \, \psi_p(\bm{\xi}(\bm{q})) \
\end{align}
where $u_p=u(\bm{q}(\bm{\xi}_p))$ are the nodal coefficients (field values at the grid points) and $\psi_p(\bm{\xi})=\ell_{p_1}(\xi) \, \ell_{p_2}(\eta)$ are the Lagrange interpolating polynomials with respect to the collocation points $\bm{\xi}_p$. Details can be found in \ccites{Vu:2024,Vu:2021}. The polynomial order can be chosen in each element and dimension individually by varying $N_\xi$ and $N_\eta$. Our $p$ refinement strategy in this work is to begin with $4 \times 4$ grid points per element initially, and then increase the polynomial order in all elements \emph{that do not contain the puncture} by 1 in both dimensions in each refinement step. The expectation is that the spectral expansion, \cref{eq:spectral_sum}, converges exponentially fast for smooth fields. In addition, note that the spectral expansion imposes regularity on the space of discrete representations of the fields. Therefore, numerical solutions are always regular by construction.

\subsubsection{DG residuals}

Finally, we discretize the PDEs, $\triangle_m\Psi_m = S_m$. To this end, we define nodal DG residuals on the grid by projecting the equations onto the same polynomial basis used to expand fields in each element, $\psi_p(\bm{\xi})$.
Projecting the first-order flux divergence form of the equations, \cref{eq:first-order}, and following \ccite{Vu:2024}, gives the DG residuals in strong form,
\begin{align}\label{eq:dgres_strong}
\begin{split}
  &-(\bas_p, \partial_i \dgF_m^{i})_{\Omega_k}
  + (\bas_p, \beta_m \Psi_m + \gamma^i_m\partial_i \Psi_m)_{\Omega_k} \\
  &\qquad+ \big(\partial_i \bas_p, \dgFA_m^{ij} n_j (\Psi_m^* - \Psi_m)\big)_{\partial \Omega_k} \\
  &\qquad- \big(\bas_p, (n_i \dgF_m^i)^* - n_i \dgF_m^i\big)_{\partial \Omega_k}
  = (\bas_p, S_m)_{\Omega_k}
\end{split}
\end{align}
where $\dgFA_m^{ij}$ is a (diagonal) matrix with coefficients defined by the linear form of the flux, $\dgF_m^{i}\equiv\dgFA_m^{ij}\partial_j \Psi_m$, in \cref{eq:flux}, $n_i$ is the normal to the element boundary, and the inner products are defined as integrals over each element or element boundary,
\begin{subequations}\label{eq:dgproj}
\begin{align}
  (\phi, \pi)_{\Omega_k} \coloneqq{}& \int_{[-1,1]^2} \phi(\bm{\xi}) \pi(\bm{\xi}) \, \jac \dd{^2\xi} \\
  (\phi, \pi)_{\partial\Omega_k} \defeq{}& \int_{[-1,1]} \phi(\bm{\xi}) \pi(\bm{\xi}) \, \surf{\jac} \dd{\xi}
\end{align}
\end{subequations}
with the Jacobian $\jac$ or surface Jacobian $\surf{\jac}$ of the coordinate map to logical coordinates, $\bm{q}(\bm{\xi})$.
At element interfaces, neighboring elements are coupled through the numerical flux, $\Psi_m^*$ and $(n_i \dgF_m^i)^*$. For this coupling we choose the internal penalty flux,~\cite{Vu:2024}
\begin{subequations}\label{eq:numflux}
\begin{align}
\Psi_m^* &= \frac{1}{2} \left(\Psi_m^\mathrm{int} + \Psi_m^\mathrm{ext}\right) \\
(n_i \dgF_m^i)^* &= \begin{aligned}[t]
  &\frac{1}{2} n_i \left(\dgFi{\mathrm{int}} + \dgFi{\mathrm{ext}} \right) \\
    &- \sigma \, n_i \dgFA_m^{ij} n_j (\Psi_m^\mathrm{int} - \Psi_m^\mathrm{ext})
    \text{,}
  \end{aligned}
\end{align}
\end{subequations}
where ``int'' denotes the quantities on the interior side of the element, and ``ext'' denotes the quantities on the exterior side that were sent from the neighboring element.
The factor~$\sigma$ is the penalty factor as defined in \cite{Vu:2024}, with penalty parameter $C=1.5$ chosen in this work.

\subsubsection{Imposing jump conditions}\label{sec:jump_cond}

With the DG numerical flux we also impose the jump conditions across the worldtube boundary, \cref{eq:jump_worldtube}, and between the $v$, $t$, and $u$ domains, \cref{eq:jump_vt,eq:jump_tu}. To this end, we modify the exterior quantities $\Psi_m^\mathrm{ext}$ and $(n_i\dgF_m^i)_\mathrm{ext}$ once they were received by an element. To impose worldtube jumps, \cref{eq:jump_worldtube}, we modify
\begin{align}
    \Psi_m^\mathrm{ext} &\to \Psi_m^\mathrm{ext} \pm \Psi_m^\mathcal{P} \\
    (n_i\dgF_m^i)_\mathrm{ext} &\to (n_i\dgF_m^i)_\mathrm{ext} \mp n_i\dgFA_m^{ij}\partial_j\Psi_m^\mathcal{P}
\end{align}
on elements that receive data from across the worldtube boundary.
Similarly, on $vtu$ domain interfaces, \cref{eq:jump_vt,eq:jump_tu}, we modify
\begin{equation}
    (n_i\dgF_m^i)_\mathrm{ext} \to (n_i\dgF_m^i)_\mathrm{ext} - im\Omega\,\{\!\{\Psi_m\}\!\}
    \text{,}
\end{equation}
where $\{\!\{\Psi_m\}\!\}=\frac{1}{2} (\Psi_m^\mathrm{int} + \Psi_m^\mathrm{ext})$ is the average across the boundary (note that the change in sign on either side of the boundary is absorbed by the opposite face normal, $n_i$).

\subsubsection{Imposing boundary conditions}\label{sec:bc_impl}

On external boundaries we impose boundary conditions through the DG numerical flux by a choice of $\Psi_m^\mathrm{ext}$ and $(n_i\dgF_m^i)_\mathrm{ext}$. Note that on boundaries where the perpendicular flux vanishes, $n_i\dgFA^i_m=0$, neither of the two boundary terms contributes, as $\Psi_m^\mathrm{ext}$ is ignored and $(n_i\dgF_m^i)_\mathrm{ext}=0$. This occurs on the horizon boundary and on angular boundaries. On the last remaining boundary at finite outer radius, $R=10^4$, we impose the second-order Bayliss-Turkel condition, \cref{eq:bayliss_turkel_bc}, by the choice
\begin{align}
    &(n_i\dgF^i)_\mathrm{N} = -\frac{\Delta}{r^2+a^2}\frac{\beta_m}{\gamma_m^r} \Psi_m
\end{align}
and $n_i\dgFi{\mathrm{ext}} = 2 \, (n_i\dgF^i)_\mathrm{N} - n_i\dgFi{\mathrm{int}}$, $\Psi_m^\mathrm{ext}=\Psi_m^\mathrm{int}$ as detailed in \ccite{Vu:2024}.

\subsubsection{Matrix representation}

\begin{figure}
    \centering
    \includegraphics[width=0.5\textwidth,clip,trim=27cm 17.5cm 31.5cm 7.7cm]{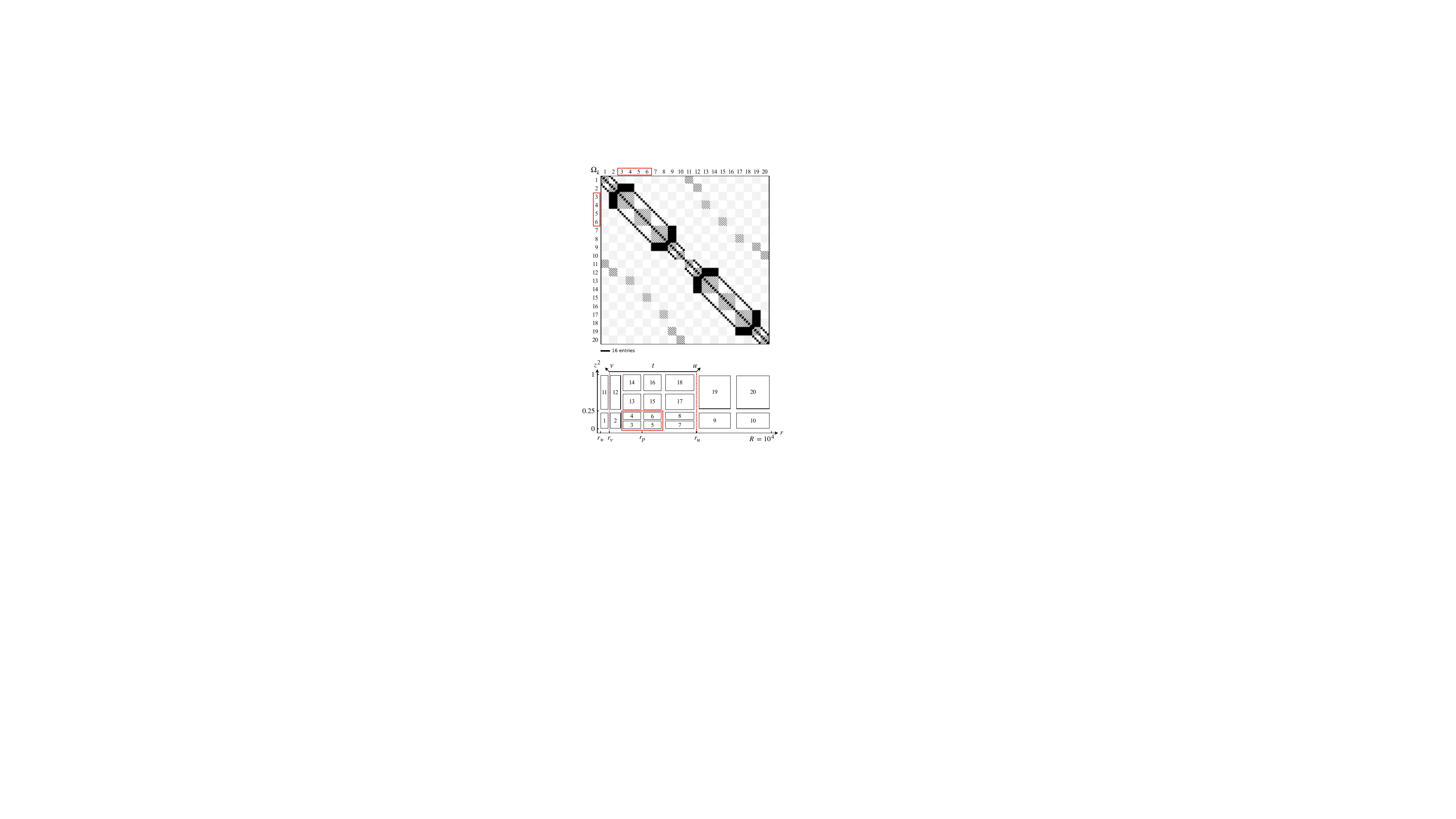}
    \caption{\textit{Top:} Matrix representation of the DG operator at the coarsest grid level (with equatorial symmetry). \textit{Bottom:} Numbering of DG elements in the domain corresponding to rows and columns in the matrix. Each element has $4\times 4$ grid points in this example. Jump conditions are imposed at the worldtube boundary (red rectangle) and between the $v$, $t$, and~$u$ domains (red dotted lines).}
    \label{fig:matrix}
\end{figure}

\begin{figure*}
    \centering
    \includegraphics[width=\textwidth,clip,trim=19.5cm 15.5cm 22.5cm 13.6cm]{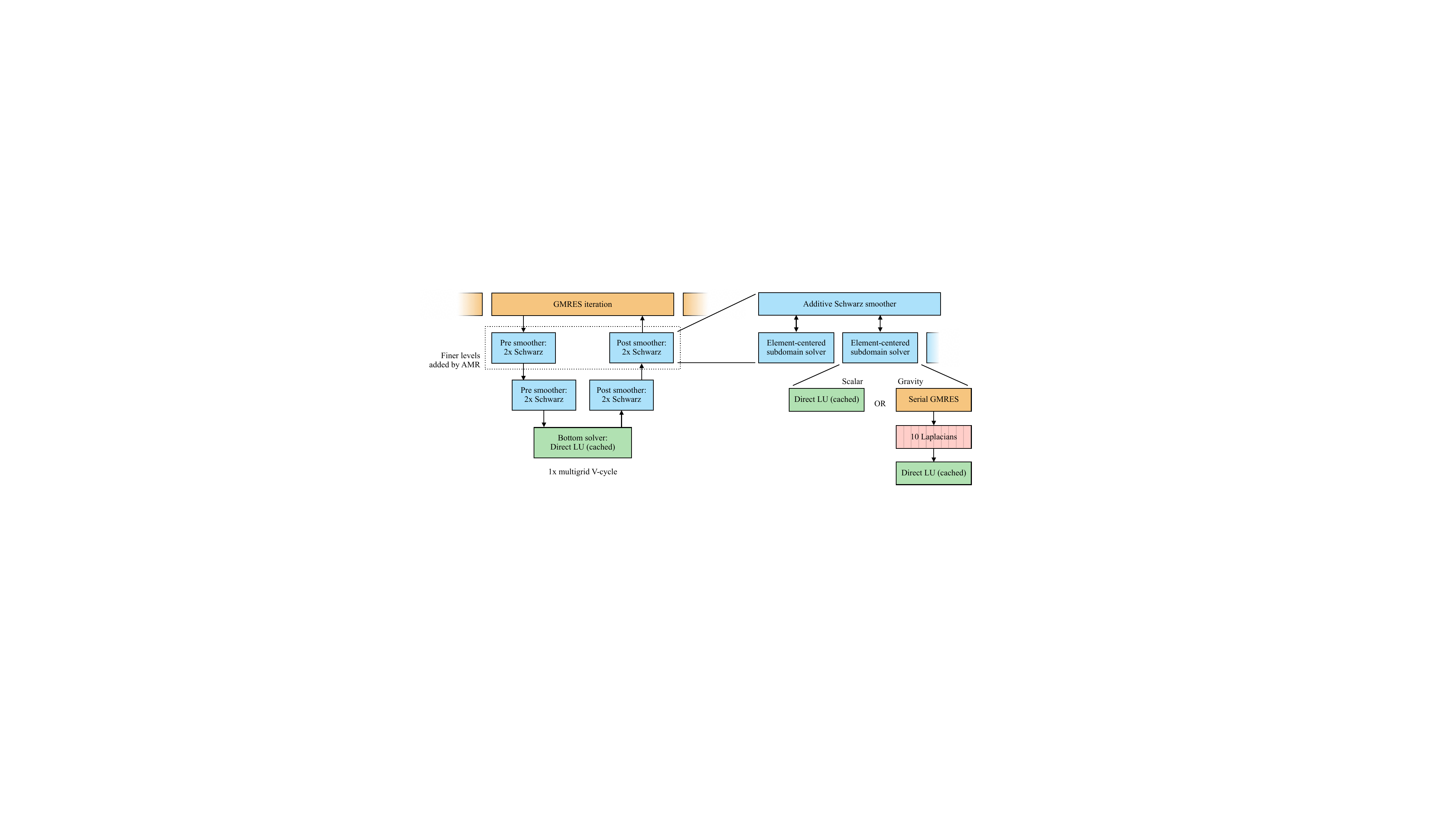}
    \caption{Preconditioning of the iterative linear solver. Each GMRES iteration is preconditioned with one multigrid V cycle, built by AMR. The coarsest level is solved directly (bottom solver). Finer levels run two additive Schwarz smoothing iterations, which solve many element-centered subdomains in parallel. Subdomains are solved directly for scalar problems, but can also run another iterative GMRES for larger (gravity) problems. In this case, the GMRES is preconditioned by approximating each component of the equations by a Laplacian, which is assembled and solved directly.}
    \label{fig:precond}
\end{figure*}

To obtain a discrete representation of the DG residuals, we evaluate the integrals in \cref{eq:dgres_strong} with Gauss quadrature on the Legendre-Gauss collocation points of each element~\cite{Vu:2021,Vu:2024}. This step reduces \cref{eq:dgres_strong} to
\begin{align}\label{eq:dgres_strong_matrix}
\begin{split}
  &-\dgMD_i \cdot \dgF_m^i
  + \dgM \cdot \beta_m \Psi_m + \gamma_m^i \dgMD_i \cdot \Psi_m  \\
  &\qquad+ \dgMLD^T_i \cdot \dgFA_m^{ij} n_j (\Psi_m^* - \Psi_m) \\
  &\qquad- \dgML \cdot \big((n_i \dgF_m^i)^* - n_i \dgF_m^i\big)
  = \dgM \cdot S_m
  \text{,}
\end{split}
\end{align}
where the symbol $\cdot$ denotes matrix-vector multiplication over the grid points of an element. The matrices are
\begin{align}
  &\dgM_{pq} = \delta_{pq} \atp{\jac}_p \prod_{\hat{\jmath}=1}^2 w_{p_{\hat{\jmath}}} \\
  &\dgMD_{i,pq} = \dgD_{\hat{\imath},pq} \atp{\jac}_q \atp{\invjac^{\hat{\imath}}_i}_q \prod_{\hat{\jmath}=1}^2 w_{p_{\hat{\jmath}}} \\
  &\dgML_{p\bar{q}} = I_{p\bar{q}} \atp{\surf{\jac}}_{\bar{q}} w_{\bar{q}_{\hat{\jmath}}} \\
  &\dgMLD^T_{i,p\bar{q}} = \dgD^T_{\hat{\imath},pq} I_{q\bar{q}} \atp{\surf{\jac}}_{\bar{q}} \atp{\invjac^{\hat{\imath}}_i}_{\bar{q}} w_{\bar{q}_{\hat{\jmath}}}
  \text{,}
\end{align}
where $w_{p_{\hat{\jmath}}}$ are the Gauss quadrature weights
in logical dimension $\hat{\jmath}$ of the element,
\begin{equation}\label{eq:diffmat}
\dgD_{\hat{\imath},pq}=\lagr^\prime_{q_{\hat{\imath}}}(\xi_{p_{\hat{\imath}}})\prod_{\hat{\jmath}\neq \hat{\imath}}\delta_{p_{\hat{\jmath}} q_{\hat{\jmath}}}
\end{equation}
is the differentiation matrix in logical dimension $\hat{\imath}$, and
\begin{equation}
  I_{p\bar{q}}=\lagr_{p_{\hat{\imath}}}(\pm 1)\prod_{\hat{\jmath}\neq \hat{\imath}}\delta_{p_{\hat{\jmath}} \bar{q}_{\hat{\jmath}}}
\end{equation}
is the interpolation matrix from grid points in the element volume, $p$, to grid points on the element face, $\bar{q}$, on the lower ($-$) or upper ($+$) side in logical dimension $\hat{\imath}$.

On nonconforming element interfaces, where elements meet two to one ($h$ nonconforming) or have different resolution along the interface ($p$ nonconforming), additional logic is needed to compute the discrete numerical fluxes between elements ($\Psi_m^*$ and $(n_i\dgF^i_m)^*$ in \cref{eq:dgres_strong_matrix}). To this end, we define ``mortars'' between elements, which cover the full element face with the larger of the two abutting resolutions~\cite{Vu:2021}. Both neighboring elements project their interface data to the mortar, compute the numerical flux there, and restrict the result back to their face. The projection operator from the element face to the mortar is a Lagrange interpolation and the restriction is its mass-conservative adjoint, as detailed in \ccite{Vu:2021}.

Taken together, the action of these matrices on the nodal field values forms a matrix-vector product, schematically $A\cdot u$, that represents the discretized equations, \cref{eq:dgres_strong_matrix}. \Cref{fig:matrix} shows the matrix representation of the operator $A$. It can be inverted in the next step to obtain the solution. However, note that this matrix is never assembled explicitly in the solver, but only used to compute the matrix-vector product. Therefore, only constituents such as the one-dimensional logical differentiation matrices, $\lagr^\prime_{q_{\hat{\imath}}}(\xi_{p_{\hat{\imath}}})$, are computed explicitly and cached globally in the simulation. This matrix-free scheme allows one to scale computations effectively to high resolution because no large matrices must be assembled and stored, but requires iterative linear solvers as discussed in the next section.

\subsection{Iterative linear solver}\label{sec:linsolv}

Once we have a discretized matrix equation, $A\cdot u=b$, it is the job of the linear solver to invert it for the nodal field values $u$. We use an iterative Krylov method, specifically a complex-valued generalized minimal residual (GMRES) algorithm, with the stack of preconditioners detailed in \ccite{Vu_2022}, sketched in \cref{fig:precond}, and summarized here.

Each GMRES iteration is preconditioned with one multigrid V cycle, where the multigrid hierarchy is built by $h$-coarsening the domain up to the initial level (the irreducible block structure). At the coarsest level, the multigrid solver explicitly constructs the matrix $A^\text{bottom}$ column by column and inverts it directly using the \codename{Eigen} sparse lower-upper (LU) decomposition routines (bottom solver). This helps provide an accurate large-scale baseline solution. The LU decomposition of $A^\text{bottom}$ is cached and reused in all multigrid V cycles across AMR iterations, since the bottom grid remains the same even when AMR adds finer levels.

On all finer multigrid levels, we use an additive Schwarz smoother. It works by constructing element-centered subdomains, which cover the DG element and a few points overlap into each face neighbor, assuming all data outside the subdomain is zero. It explicitly constructs the subdomain matrix $A^\text{subdomain}$ column by column, inverts it directly, and caches the inverse. Now, one Schwarz-smoothing pass on a multigrid level involves solving many independent subdomain problems (one per DG element on that level) by applying the cached inverse, and then combining the overlapping subdomain solutions in a weighted sum. We run two Schwarz-smoothing passes per multigrid level. Also here, the matrix inverses can remain cached across AMR iterations, so we only have to construct new subdomain matrices when AMR adds a finer level. Constructing subdomain matrices is typically the most expensive operation in the elliptic solver, but can be parallelized over DG elements.

For problems where each subdomain matrix is still too large to assemble and solve explicitly, as is likely the case for gravitational self-force problems that involve ten degrees of freedom per grid point instead of one, we insert another layer of preconditioning: we solve each subdomain problem iteratively with a GMRES algorithm and precondition it by approximating the subdomain problem as ten decoupled Poisson equations. These Poisson operators we can now construct explicitly on the subdomain and invert directly as before.

This technology stack has been developed for binary black hole initial data problems and has proven to work well for self-force calculations as well with a few advancements: (1) We added support for complex-valued fields in the GMRES algorithms. Technically, treating complex-valued fields as two real-valued fields works as well, but the complex-valued GMRES converges in fewer iterations as complex arithmetic is built into the algorithm and restricts the solution space. (2) We improved convergence of complex Helmholtz-type problems such as our elliptic self-force problem. Helmholtz-type problems are challenging because noise in the numerical solution can excite oscillations, rendering preconditioners ineffective. To combat this effect, we add a small complex shift on the order of $m^2\Omega^2$ to the subdomain matrices that dampens spurious oscillations in the Schwarz smoother~\cite{Erlangga2006,vanGijzen2007}. A commonly used alternative that we have not explored is deflation (projecting out known oscillating modes)~\cite{Dwarka2020}. We found that the complex shift is essential to solve oscillating problems in $t$-slicing, but can be reduced or disabled altogether when hyperboloidal or null slicing is used as the solution loses its global wavelike behavior.

\section{Results}\label{sec:results}

\begin{figure}
    \centering
    \includegraphics[width=0.5\textwidth]{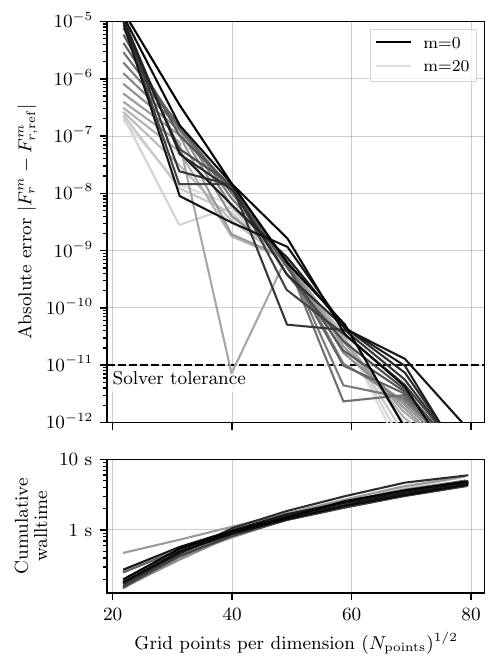}
    \caption{Convergence of the DG method for modes up to $m=20$, exemplary for the $a=0.5 M$, $r_0=10 M$ configuration. The convergence is exponential despite the nonsmooth solution on the grid under our $hp$-AMR scheme. Each mode solves in a few seconds on an Apple M2 Pro chip.}
    \label{fig:convergence}
\end{figure}

\begin{figure*}
    \centering
    \includegraphics[width=\textwidth,clip,trim=0 6.1cm 0 3.3cm]{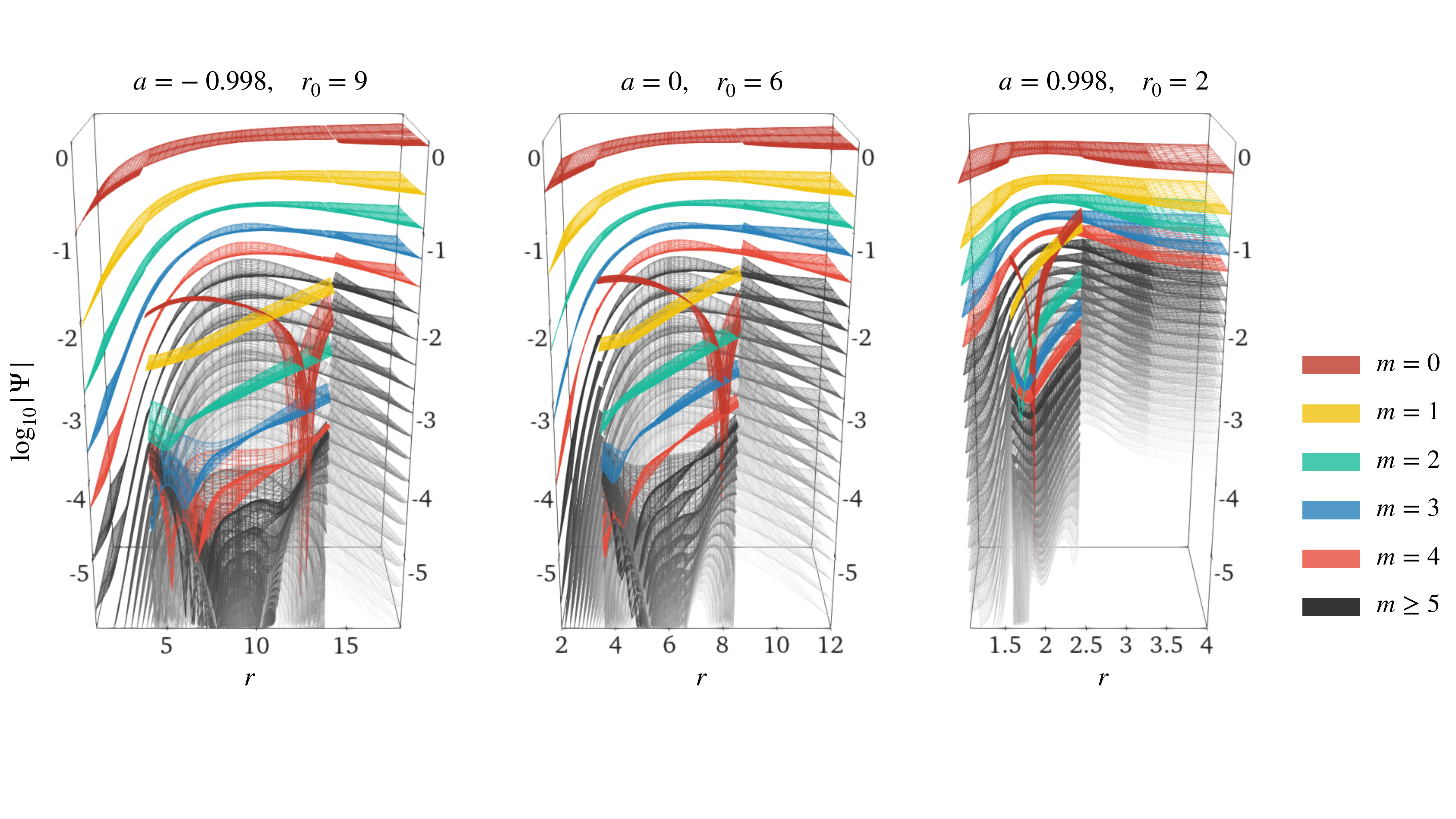}
    \caption{$m$-mode solutions for a scalar charge orbiting a black hole at the ISCO for three different black hole spins. Each plane shows an $m$-mode of the scalar field, $\Psi_m$, up to $m=20$, with the first few modes highlighted in color. Plotted is their complex magnitude. Within the regularized region, the regular field $\Psi^\mathcal{R}_m=\Psi_m-\Psi^\mathcal{P}_m$ is plotted. The computational domain extends over $z^2\in [0, 1]$ (equatorial symmetry) and to $R=10^4 M$ toward the right, but the plot is clipped at $r=2\, r_0$.}
    \label{fig:modes}
\end{figure*}

\begin{figure}
    \centering
    \includegraphics[width=0.5\textwidth]{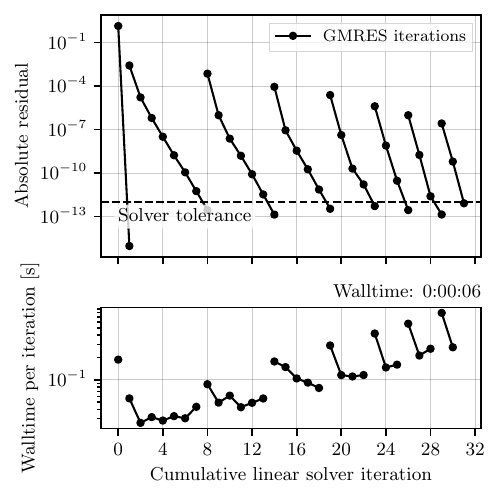}
    \caption{Convergence of the preconditioned linear solver, exemplary for the $m=2$ mode of the $a=0.5 M$, $r_0=10 M$ configuration. The lines in the top panel, from left to right, show the GMRES iterations needed to solve consecutive AMR levels. Each solve converges in less than 10 preconditioned iterations. The first AMR level converges immediately because the multigrid bottom solver solves the problem directly. The bottom panel shows the walltime for each iteration. The first iteration on each new AMR level is slower than the following because it builds and caches subdomain matrices.}
    \label{fig:linsolv}
\end{figure}

\begin{figure*}
    \centering
    \includegraphics[width=\textwidth]{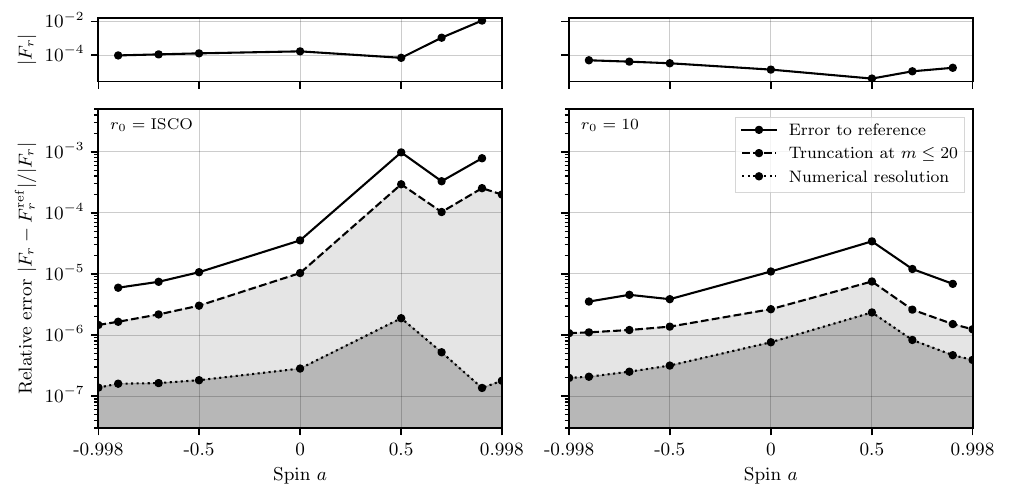}
    \caption{Error budget in the extracted self-force. Upper panels show the value of the self-force and lower panels the relative error. The numerical resolution floor is below the error to the reference solutions from \ccite{Warburton:2010eq}, indicating that accuracy is not limited by numerical resolutions. Instead, the error is dominated by truncation error in the sum over $m$-modes.}
    \label{fig:error_budget}
\end{figure*}

We demonstrate the capability of our new method by solving scalar self-force problems in Kerr spacetime at a wide range of orbital radii. We explore spins up to magnitude $|a|/M=0.998$ (Thorne limit) with prograde and retrograde orbits, both at $r_0=r_\mathrm{ISCO}$ and at $r_0=10 M$. We extract the radial self-force contributions $F_r^m$ from each $m$-mode solution up to $m=20$ following Eq.~(3.10) in \ccite{Osburn_2022}, sum over $m$-modes, and add regularization parameters to obtain $F_r$ following \ccite{Heffernan_2014} (see also \cref{sec:smoothness}). Then, we compute the relative error, $|F_r - F_r^\mathrm{ref}| / F_r$, where $F_r^\mathrm{ref}$ is an accurate reference solution taken from Warburton and Barack~\cite{Warburton:2010eq}. Input-file configurations to reproduce our results with the open-source \spectre{} code~\cite{spectrecode} are available as supplemental material~\cite{suppl}.

\Cref{fig:convergence} demonstrates that we achieve exponential convergence with our DG scheme despite the nonsmooth nature of the solution. We plot the absolute error for each individual $m$-mode contribution, $|F_r^m - F_{r,\mathrm{ref}}^m|$, up to $m=20$ as a function of the number of grid points across a sequence of AMR iterations. As discussed in \cref{sec:refinement}, in each AMR iteration we split elements that contain the puncture in two in both dimensions ($h$-refinement) and increase the polynomial order everywhere else by 1 ($p$-refinement). As reference solution for each $m$-mode, $F_{r,\mathrm{ref}}^m$, we use the result from the highest-resolution AMR iteration here. The error decreases exponentially with the number of grid points until it reaches the tolerances set for the GMRES algorithm and for AMR.\footnote{Specifically, we expect the error to decrease exponentially with $(N_\text{points})^{1/d}$, where $d>2$, based on earlier explorations of the puncture initial data problem under a similar $hp$-AMR strategy, see \ccite{Vu:2024}. We have not yet performed a thorough analysis of the expected power $d$ for this problem and therefore plot the error over the number of grid points per dimension for an intuitive measure of resolution.} For more generic orbits where the source position is averaged over a line (eccentric orbits or spherical orbits) or over an extended region (generic orbits), a more careful treatment may be required. Nevertheless, since exponential convergence relies solely on our ability to $h$-refine nonsmooth points, we are optimistic that this convergence carries over to more generic cases provided they are handled with sufficient care.

The bottom panel of \cref{fig:convergence} shows the cumulative walltime for each $m$-mode solve. The full AMR sequence takes a few seconds per $m$-mode to complete on an Apple M2 Pro chip. Each $m$-mode is independent, so they can be computed in parallel on a computing cluster.

\Cref{fig:modes} gives an overview of the $m$-mode solutions we obtain with our method. All $m$-modes up to $m=20$ are shown on a log scale, demonstrating the decay in power over mode number. Note that regular zero crossings in the solution of the regularized field within the worldtube display as sharp drops on the chosen log scale, particularly for the $m=0$ mode. Also visible are features in the solution within the worldtube that are resolved by our adaptive grid.

\Cref{fig:linsolv} demonstrates the effectiveness of our preconditioned iterative linear solver. It shows the convergence of GMRES iterations across consecutive AMR levels for one exemplary mode out of the $m$-modes plotted in \cref{fig:convergence}. We find that the preconditioners, as detailed in \cref{sec:linsolv}, allow the GMRES algorithm to converge rapidly in less than ten iterations per AMR level. In particular, the multigrid algorithm achieves scale independence, meaning that the number of GMRES iterations does not increase when AMR levels are added, though each multigrid V cycle increases in computational cost. The computational cost is dominated by Schwarz subdomain solves on each multigrid level, which can parallelize over elements in our additive Schwarz algorithm. We make substantial use of caching to avoid recomputing LU decompositions of matrices that remain constant throughout the AMR sequence (subdomain matrices in multigrid bottom solver).

Finally, \cref{fig:error_budget} shows the accuracy of the full radial self-force, $F_r$, up to spin $|a|/M=0.998$ (prograde and retrograde), both at the ISCO and at $r_0=10 M$. We use the $F_{r,[4]}$ regularization parameter from \ccite{Heffernan_2014} to accelerate the sum over $m$-modes in our solution. We measure the error due to numerical resolution by the difference between the highest and second-highest AMR level. We compute the true error by the difference to the accurate reference solution from \ccite{Warburton:2010eq} and we estimate the error due to finite-$m$ truncation by the difference between using all $m\leq 20$ modes and omitting the highest mode, $m=20$.
We find that the error due to numerical resolution is pushed well below the finite-$m$ truncation error, as a consequence of the exponential convergence of our method.
We also find that our finite-$m$ truncation error models the true error well up to a constant offset. Fits for higher-order regularization parameters have shown to increase accuracy in previous work and could be applied to our solutions as well~\cite{Heffernan_2014}.
The accuracy can be further increased by using a higher-order puncture or more $m$-modes.
Note that the relative errors in \cref{fig:error_budget} peak around $a=0.5$ because of a zero crossing in $F_r$, which causes a decrease in $|F_r|$ (see top panels in \cref{fig:error_budget}). This change of sign of the radial self-force is documented in \ccite{Warburton:2010eq}, Sec.~V.D.

\section{Conclusions and future directions}\label{sec:conclusion}

We present a new set of computational methods to solve self-force problems using techniques gleaned from numerical relativity. We formulate the scalar self-force problem as an elliptic $m$-mode problem with multiple advancements compared to previous elliptic formulations, including piecewise-constant null slicing (\vtu{} slicing) with horizon-penetrating coordinates. Then, we use the elliptic solver from the open-source numerical relativity platform \spectre{}, which was originally developed to generate initial data for binary black hole simulations, to solve the elliptic problem with a discontinuous Galerkin discretization and an iterative Krylov-type linear solver with multigrid-Schwarz preconditioning. With this new method we can solve challenging scalar self-force problems, such as prograde orbits at the ISCO in Kerr spacetime with spin magnitude $|a|/M=0.998$ (Thorne limit), in a few seconds per $m$-mode (parallelizable over $m$-modes). We demonstrate exponential convergence of the method under $hp$ AMR (splitting elements at the particle and increasing spectral order everywhere else), which pushes numerical resolution error below $m$-mode truncation error across the parameter space when including 20 $m$-modes.

It is valuable to compare and contrast this work with that of Macedo et al.~\cite{Macedo_2024}, which involve a number of similarities; indeed, their work inspired many of the techniques implemented here including hyperboloidal slicing, discretizing with $z=\cos{\theta}$, and factoring out $\text{sin}^{|m|}\theta$ from the mode functions. Advancements in our work include the use of DG methods to impose jump conditions, $hp$ AMR to achieve exponential convergence despite the nonsmooth solution on the grid, and preconditioned iterative methods to solve the elliptic problem. Other aspects of \ccite{Macedo_2024} are not included in our work but arguably more elegant than some of our choices. Specifically, their delicate domain decomposition involving a locally polar coordinate map surrounding the particle is able to isolate the nonsmooth behavior along the locally radial direction and enable rapid convergence along the locally angular direction; in contrast, our AMR method $h$ refines in both radial and angular directions near the particle. This suggests both approaches could benefit from merging their most advantageous features, perhaps by collaborating to introduce the locally polar coordinate map as an option in a future \spectre{} implementation. However, we want to emphasize a certain important advantage of our approach: even with more sophisticated domain decompositions, it is likely that refining in two spatial dimensions near the particle (as our AMR method does) will be required to handle generic Kerr orbits, which justifies a long-term commitment to maintaining and improving that feature for self-force calculations.

Looking toward the future, major work is underway to generalize these methods to eccentric and/or inclined motion, as well as to the gravitational self-force problem at both first and second order, which is the true target of our work.
We expect the generality of the domain decomposition and refinement routines to perform favorably when advancing to these cases, and we expect the elliptic solver to prove capable of solving the larger Lorenz gauge metric perturbation problem (with ten degrees of freedom per grid point instead of one) as it has demonstrated the capability to solve much larger 3D numerical relativity problems. With these new computational tools at hand, we aim for the ultimate goal of calculating Kerr metric perturbations through second order in perturbation theory by solving the first-order problem at high accuracy, and then constructing and/or importing a second-order source to solve the second-order problem.

Several additional minor areas of ongoing work can improve the solver further. A smoother/higher-order puncture would produce a solution that converges faster. Full compactification of the outer wave zone would eliminate the need for the last remaining boundary condition at the outer radial boundary and would provide direct access to the waveform at future null infinity. And additional Helmholtz-type preconditioners for the iterative linear solver, such as deflation techniques~\cite{Dwarka2020}, may improve convergence.

\begin{acknowledgments}
We thank Saul Teukolsky, Rodrigo Panosso Macedo, Cillian Kelly, Christiana Pantelidou, Brad Cownden and Patrick Bourg for helpful discussions, as well
as the \spectre{} development team for the collaborative work.
We thank Niels Warburton for providing comparison self-force data.
Computations were performed on the CaltechHPC cluster at Caltech. The figures in this
article were produced with \codename{matplotlib}~\cite{matplotlib1,matplotlib2}
and \codename{ParaView}~\cite{paraview}.
NLV acknowledges support from the Swiss National Science Foundation (SNSF) Ambizione Grant
No.~PZ00-2\_232961. For the purpose of open access, a CC BY public copyright licence is applied to any author accepted manuscript (AAM) version arising from this submission.
NLV, TO and JET acknowledge support from the NASA LISA Preparatory
Science Grant No.~20-LPS20-0005.
This work was supported in part by the Sherman Fairchild Foundation at Caltech and Cornell; the National Science Foundation under Grants No.~PHY-2309211, No.~PHY-2309231, and No.~OAC-2209656 at Caltech; No.~PHY-2407742, No.~PHY-2207342, and No.~OAC-2513338 at Cornell; and NASA award No.~80NSSC26K0340 at Cornell.
TO gratefully acknowledges support from the National Science Foundation
under Grant No.~PHY-2309020 at SUNY Geneseo and from the Fulbright U.S. Scholar Program, which is sponsored by the U.S. Department of State and the Fulbright Commission in Ireland; the contents are solely the responsibility of the authors and do not necessarily represent the official views of the Fulbright Program, the U.S. Government, or the Fulbright Commission in Ireland.
This work is supported by ERC grant EMRIWaveforms (DOI: \href{https://doi.org/10.3030/101200625}{10.3030/101200625}).
SDU acknowledges support from the ERC Consolidator/UKRI Frontier Research Grant GWModels (selected by the ERC and funded by UKRI [Grant No.~EP/Y008251/1]).

\end{acknowledgments}

\bibliography{main}

\end{document}